\documentclass[review]{elsarticle}
\usepackage{lineno,hyperref}
\modulolinenumbers[5]
\usepackage[utf8]{inputenc}
\usepackage[francais,english]{babel}		                 

\usepackage{amsmath}						 
\usepackage{amssymb}						 

\usepackage{graphicx}
\usepackage{dcolumn}
\usepackage{bm}
\usepackage{multirow}
\usepackage{textcomp}
\usepackage[]{subfigure}
\usepackage{booktabs}
\usepackage{array}
\usepackage{pgf}
\usepackage{graphicx}
\usepackage{bm}
\usepackage[toc,page]{appendix}
\usepackage{hyperref}

\graphicspath{{}}

\begin{document}
\begin{frontmatter}
%
\title{A mean-field model of static recrystallization considering orientation spreads and their time-evolution}




\author[ubc,INP]{A.~Després\corref{cor1}}
	\ead{arthur.despres@alumni.ubc.ca}
\author[CAN]{M.Greenwood}
\author[ubc]{C.~W.~Sinclair}

\cortext[cor1]{Corresponding Author,}

\address[ubc]{Department of Materials Engineering, The University of British Columbia, 309-6350 Stores Road, Vancouver, Canada} 
\address[CAN]{Natural Resources Canada, CanmetMATERIALS, Hamilton, ON, L8P 0A5}
\address[INP]{Univ. Grenoble Alpes, Grenoble INP, SIMaP, F-38000, Grenoble, France}

\begin{abstract}

In this paper, we develop a mean-field model for simulating the microstructure evolution of crystalline materials during static recrystallization.  The model considers a population of individual cells (i.e. grains and subgrains) growing in a homogeneous medium representing the average microstructure properties. The average boundary properties of the individual cells and of the medium, required to compute growth rates, are estimated statistically as a function of the microstructure topology and of the distribution of crystallographic orientations. Recrystallized grains arise from the competitive growth between
cells. After a presentation of the algorithm, the model is compared to full-field simulations of recrystallization performed with a 2D Vertex model. It is shown that the mean-field model predicts accurately the evolution of boundary properties with time, as well as several recrystallization parameters including kinetics and grain orientations. The results allow one to investigate the role the orientation spread on the determination of boundary properties, the formation of recrystallized grains and recrystallization kinetics. The model can be used with
experimentally obtained inputs to investigate the relationship between deformation
and recrystallization microstructures.

\end{abstract}


\end{frontmatter}

\section{Introduction}

In most models of static and dynamic recrystallization, recrystallized grains arise from a competitive growth of subgrains or cells pre-existing in the deformed microstructure \cite{bailey_electron_1960,bailey_recrystallization_1962,weygand_nucleation_2000,holm_abnormal_2003,hurley_modelling_2003,zurob_quantitative_2006,wang_modeling_2011,favre_nucleation_2013,huang_review_2016}.
This viewpoint is shared for a large range of metals and alloys regardless of stacking fault energy  (e.g. silver and nickel \cite{bailey_recrystallization_1962}, copper \cite{bailey_recrystallization_1962,zurob_quantitative_2006,favre_nucleation_2013} , aluminium alloys \cite{hurley_modelling_2003,zurob_quantitative_2006}). On the other hand, the conditions leading to the development of recrystallized grains of particular orientation, and their incidence on the kinetics, remain diffucult to identify.


This challenge is to a great extent due to the large number of features involved in recrystallization. Deformed grains contain of the order of  10$^5$ subgrains \cite{hurley_modelling_2003}, out of which a handfull turn into recrystallized grains during annealing. State-of-the-art full-field models (e.g. phase field, Vertex dynamics, level-set) can simulate this many subgrains \cite{miesen_highly_2017,miyoshi_ultra-large-scale_2017}, but this is still insuficient to confidently predict recrystallization kinetics, grain size and crystallographic texture. As a result, the most significant applications of full-field models to recrystallization remain restricted to comparison with analytical model predictions \cite{holm_abnormal_2003,wang_modeling_2011} or to parametric studies on the role of some initial microstructure parameters \cite{weygand_nucleation_2000,suwa_phase-field_2008}.


Mean-field models are computationally more efficient than full-field models, but are limited by additional assumptions. In the early model of Bailey and Hirsch \cite{bailey_electron_1960,bailey_recrystallization_1962}, a subgrain is considered as a potential recrystallized grain when its radius exceeds the value where its inward capillary pressure is overcome by the outwards pressure induced by its neighbours. This model was extended by Zurob \emph{et al.} \cite{zurob_quantitative_2006} to predict the incubation period during which future recrystallized grains grow normally compared to the rest of the microstructure. This approach, however, misses the fact that \emph{every} growing subgrain satisfies the Bailey-Hirsch criterion \cite{bailey_electron_1960,bailey_recrystallization_1962}. Meeting the Bailey-Hirsch criterion is necessary but insufficient
for a subgrain to become a grain in the recrystallized state. In two separate publications, Humphreys \cite{humphreys_unified_1997}, and Rollett and Mullins \cite{rollett_growth_1997} proposed an approach that considers that a recrystallized grain forms when the growth rate of a subgrain relative to the average is positive. Notably, the model highlights the role of heterogeneous subgrain size and boundary properties on the onset of recrystallization. Despite a few interesting applications to experimental cases \cite{hurley_modelling_2003,razzak_simple_2012}, and comparisons to full-field simulations \cite{holm_abnormal_2003,syha_conditions_2012}, this approach remains much less popular than those relying on the Bailey-Hirsch criterion (e.g.  \cite{favre_nucleation_2013,huang_review_2016,dunlop_modelling_2007,beltran_mean_2015}).

As the microstructural heterogeneities giving rise to recrystallization develop during prior deformation, substantial efforts have also been made to simulate recrystallization from outputs of crystal plasticity models. In these cases, heterogeneities of subgrain size and disorientation have been attributed to inter-granular contrast of slip activity (estimated by Taylor factors) \cite{kestens_modeling_1996}, resolved shear stress \cite{wenk_deformation-based_1997}, and intragranular disorientation levels \cite{zecevic_modelling_2019}. These approaches generally focus on predicting the texture out of these heterogeneities while ignoring the recrystallization kinetics.



In this paper, we propose an extended mean-field model that builds on the
approaches described above. In our approach, a discrete population of subgrains evolves according to classic cellular growth laws, with a time-integration scheme implemented to update the
microstructural parameters. The recrystallized grains are identified based on a size threshold. The model extends beyond classic mean-field approaches by accounting
for the variation of subgrain properties with crystallographic orientation by tracking the moments of several boundary property distributions. As a result, recrystallization kinetics and recrystallized grain orientations are predicted together. This model is tested against full-field vertex simulations of subgrain growth and its extension to predicting experimental results is discussed. 

The paper starts by briefly introducing the methodology used for Vertex simulations. This serves to also familiarize the reader with the topology of the microstructures investigated. Next, the mean-field model is introduced. In the following sections, the ability of the mean-field model to reproduce the full-field simulations is shown, with a discussion on the strenghts, weaknesses and areas for further improvement.

\section{Full-field simulations}
%



\subsection{2D Vertex dynamics}


The 2D Vertex model simulations were performed following the methods described in \cite{weygand_vertex_1998,piekos_generalized_2008,mellbin_combined_2015}. In this model, grain and subgrain boundaries are discretized into vertices located at triple
junctions and along boundaries, and the velocities of each vertex
calculated as a function of the capillary forces exerted by its adjoining segments. Topological transformations account for the removal of boundaries when two
triple junctions meet \cite{weygand_vertex_1998}, when cells become smaller than a critical size \cite{weygand_vertex_1998}, or
when contacts between colliding boundaries occur \cite{piekos_generalized_2008,mellbin_combined_2015}. A single empirical coefficient controls the triggering of these transformations which, if set small enough, does not influence the results \cite{weygand_vertex_1998,piekos_generalized_2008,mellbin_combined_2015}.

If differences in volumetric energy across boundaries are neglected, the evolution of the boundary network is controlled by the microstructure topology, the boundary energies (inducing capillary forces) and their mobilities. As in previous work on recrystallization \cite{humphreys_unified_1997,holm_abnormal_2003,huang_subgrain_2000}, the boundary energy and mobility are assumed to be functions of the boundary disorientation\footnote{Following the standard terminology \cite{engler_introduction_2009}, a misorientation is defined as a rotation  (described by an axis and an angle) that transforms one crystalline orientation into another. The disorientation is the misorientation having the smallest rotation angle out of all misorientations allowed by the crystal symmetry.} angle $\theta$. For the purposes of this study the boundary energy $\gamma\left(\theta\right)$ is taken to obey the Read-Shockley equation \cite{read_dislocation_1950}:


\begin{equation}
\gamma\left(\theta\right)=
     \begin{cases} 
    \gamma_{c}\frac{\theta}{\theta_{c}}\left(1- \ln\frac{\theta}{\theta_{c}}  \right) & \text{if } \theta \leq \theta_c \\
    \gamma_c & \text{if } \theta > \theta_c
  \end{cases}
\label{eq:ReadSchockley}
\end{equation}

Where $\gamma_c$ is a constant, $\theta_c$ is a cut-off angle set to 15$^\circ$ to simulate a high angle boundary. The boundary mobility $\mu\left(\theta\right)$ was set to follow the empirical relation \cite{humphreys_unified_1997,huang_subgrain_2000}:

\begin{equation}
\mu \left(\theta\right)=\mu_{c}\left(1- e^{-B\left(\frac{\theta}{\theta_{c}}\right)^{\eta}}  \right)
\label{eq:Huang}
\end{equation}

Where $\mu_c$ is a constant, $B=5$ and $\eta=4$, following classic work on aluminium alloys \cite{humphreys_unified_1997,huang_subgrain_2000}.

\subsection{Microstructure construction}
%


The starting subgrain microstructures were constructed by Voronoi tesselation with periodic boundary conditions. Each Voronoi cell constitutes a subgrain. A first relaxation of the microstructure was performed by setting all boundary mobilities and energies equal, until the subgrain radii reached the self-similar distribution associated with normal growth (i.e. for 2 dimensional microstructures a Rayleigh distribution with maximum around 2 times the mean radius \cite{weygand_vertex_1998,louat_theory_1974,srolovitz_computer_1984}). The results presented in this paper were obtained from averages of 6 simulations performed with  2.5$\times10^4$ subgrains in this `as-relaxed' state.

\begin{figure}[htbp]
    \centering
    \subfigure[]{
    \includegraphics[width=0.4\textwidth]{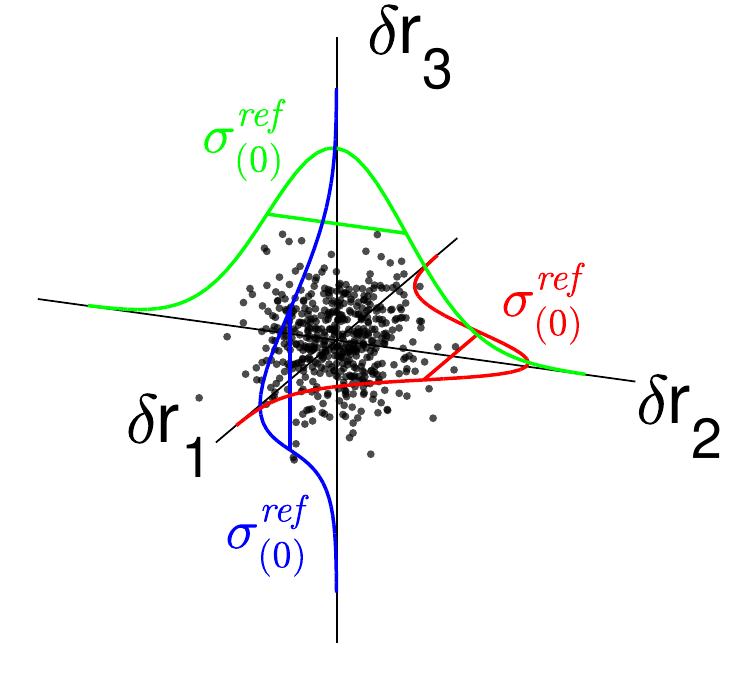}}
        \subfigure[]{
    \includegraphics[width=0.5\textwidth]{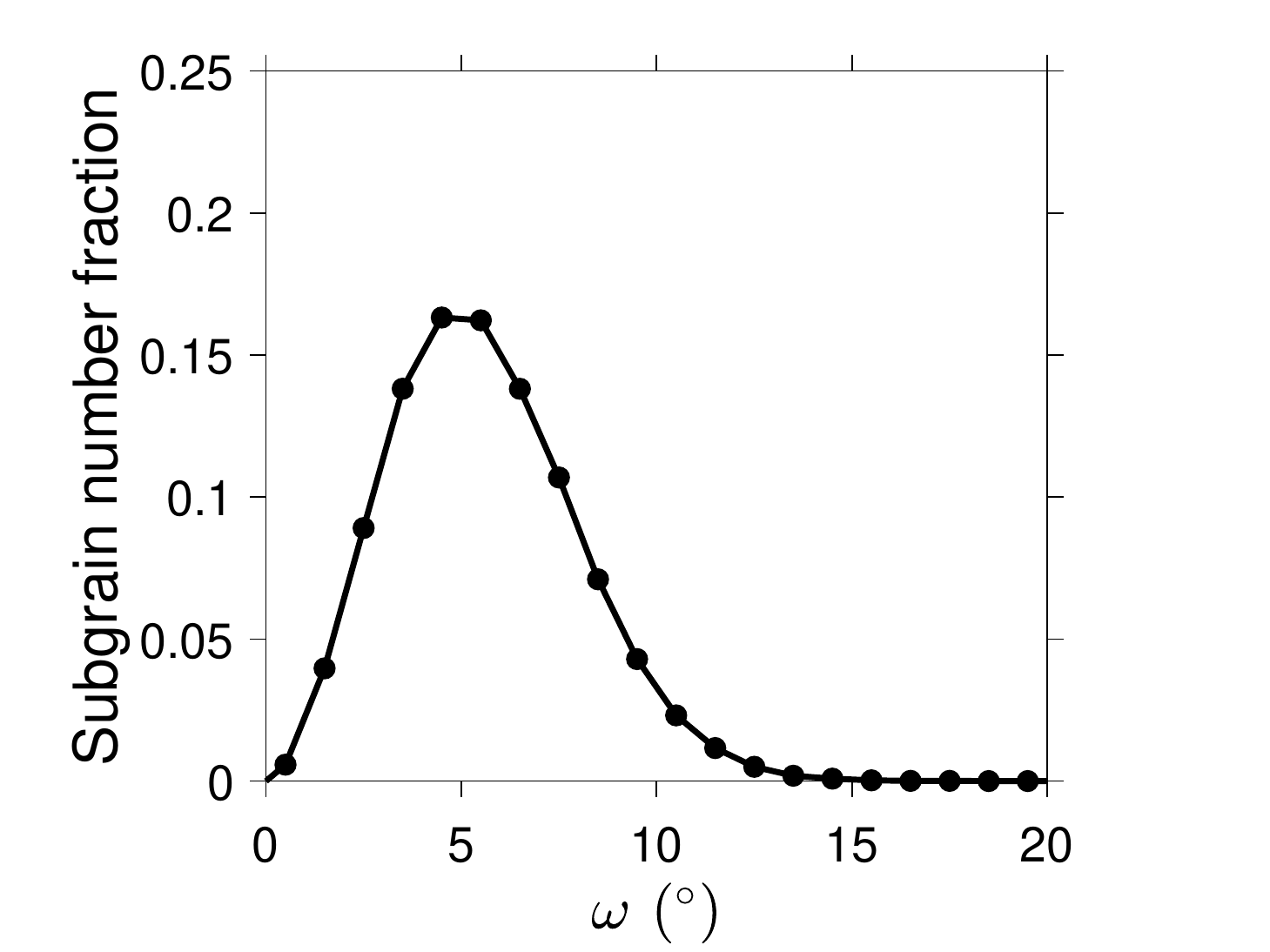}}
    \caption{a) In dots, an orientation spread represented in the quaternion vector space. The trivariate normal distribution is also represented, with $\sigma^{\textit{ref}}_{\left(0\right)}$ the standard deviation in the three directions. b) The distribution of disorientation angles $\omega$ for an isotropic spread $\sigma^{\textit{ref}}_{\left(0\right)}$=3.5$^\circ$.}
    \label{fig:spread-ex}
\end{figure}
\begin{figure}[htbp]
    \centering
    \subfigure[]{\includegraphics[width=0.49\textwidth]{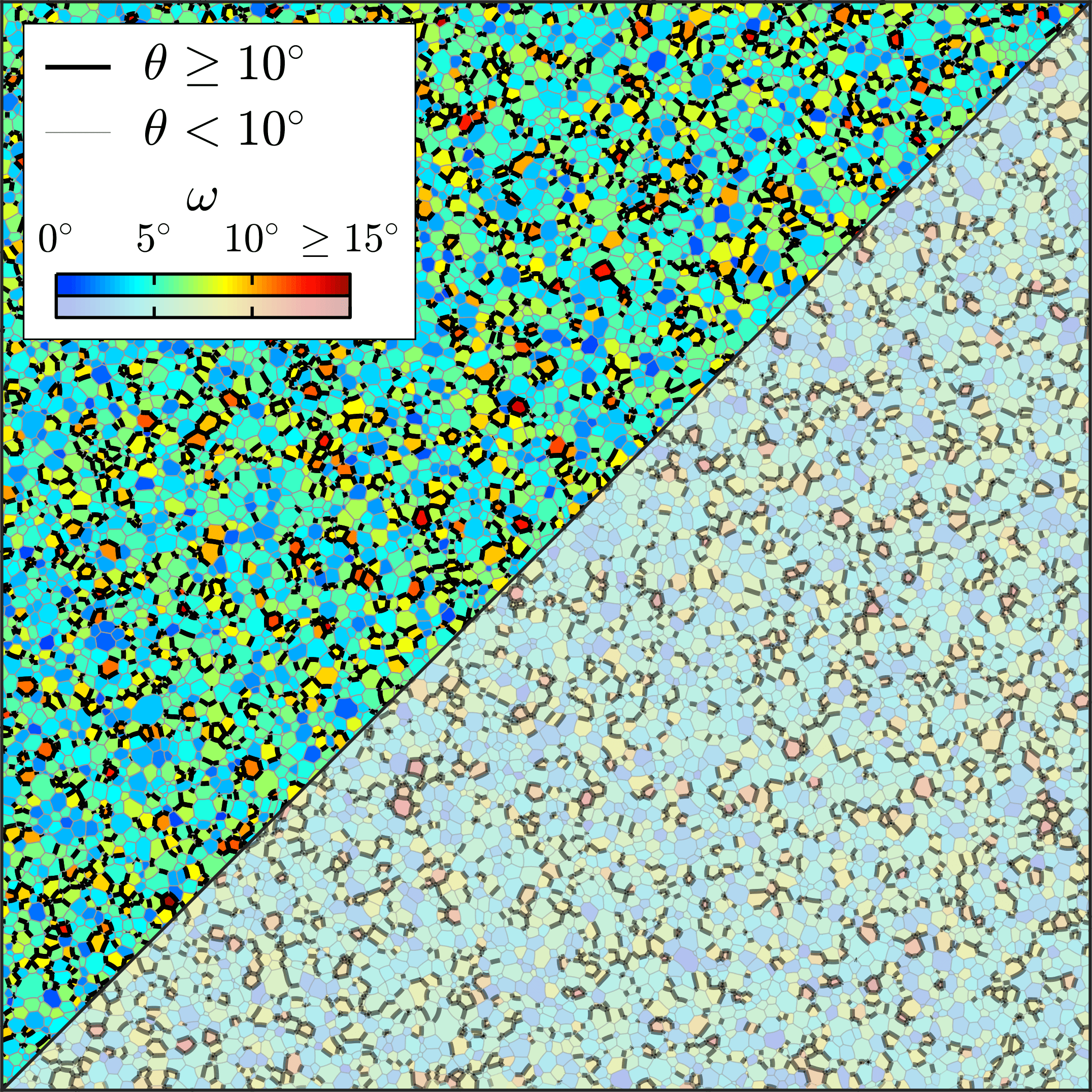}}
        \subfigure[]{\includegraphics[width=0.49\textwidth]{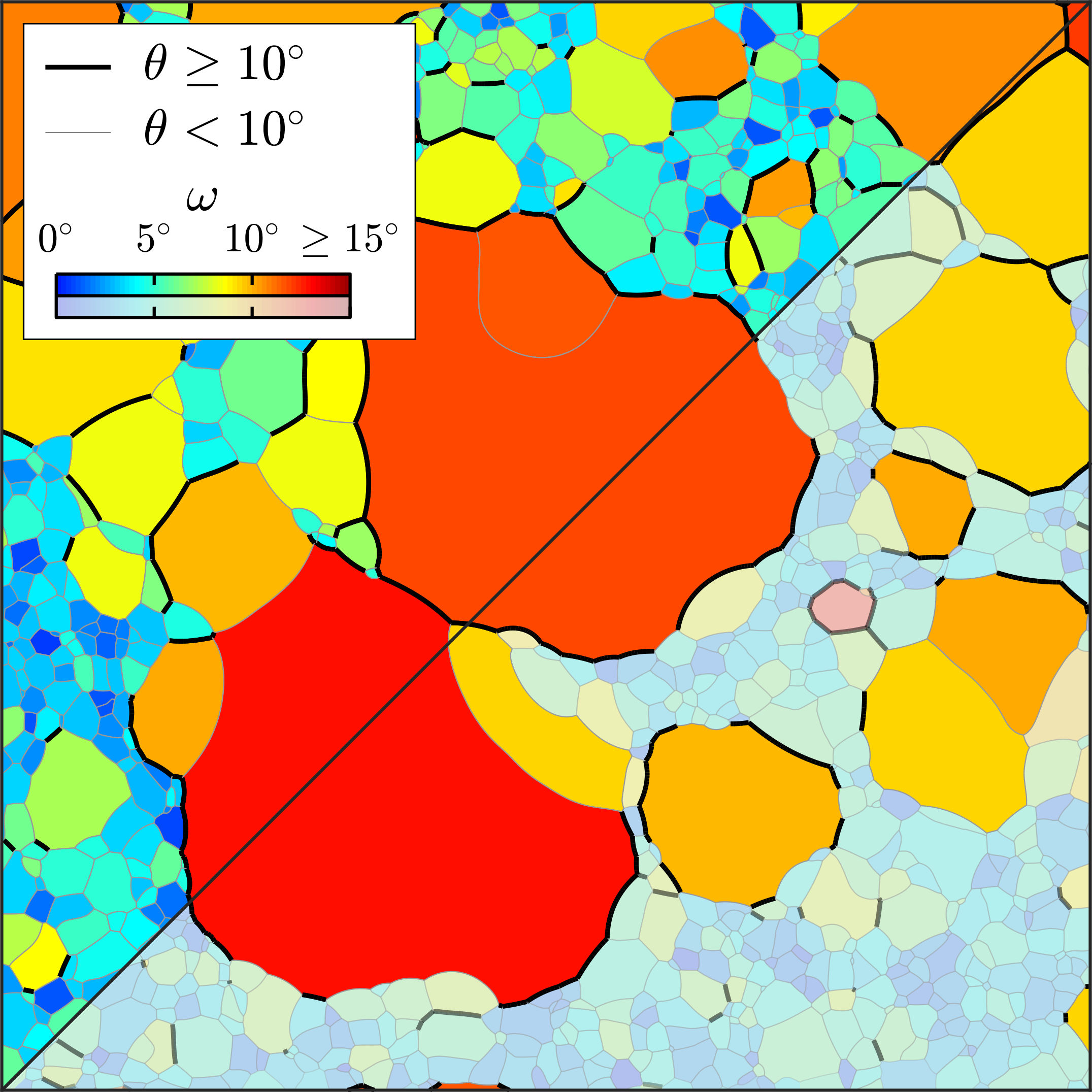}}
     
    
    \caption{a) initial microstructure simulated with Vertex dynamics for a spread $\sigma^\mathit{ref}$=3.5$^\circ$. b) same microstructure at 50\% recrystallization. In the bottom right corner, recrystallized grains are highlighted in brighter colors. Only half of the microstructure appears on the figures.}
    \label{fig:vertex-micro}
\end{figure}

Each subgrain in the `as-relaxed' microstructure was assigned a crystallographic
orientation assuming cubic symmetry and no spatial correlation. Orientations are described here by their disorientation relative to an arbitrary reference orientation, and denoted in quaternion vector part $\delta\bm{r}^{ref}=\left(r_1,r_2,r_3\right)^\mathit{ref} sin\left(\omega/2\right)$, with $\left(r_1,r_2,r_3\right)^\mathit{ref}$ the disorientation axis and $\omega$ the disorientation angle. This notation is commonly used to describe orientations spread around the mean orientation of deformed grains \cite{glez_orientation_2001,pantleon_retrieving_2005,zecevic_modelling_2019}. The initial orientations are drawn from a trivariate normal distribution along the principal directions $\delta\bm{r}_1$, $\delta\bm{r}_2$, $\delta\bm{r}_3$ of the reference frame. The distribution is centered on $\left(0,0,0\right)$ and controlled through an isotropic standard deviation $\sigma_{\left(0\right)}^{\textit{ref}}$, set identical in the three directions. It remains a trivariate normal distribution as long as the largest disorientation vectors (imposed by $\sigma_{\left(0\right)}^{\textit{ref}}$) do not exceed the bounds of the orientation space set by the symmetry of the crystal. By convention, we consider only positive disorientation angles with the vector direction carried by the sign of the rotation axis. A representation of the reference disorientation distribution is shown in \autoref{fig:spread-ex}. An example of a relaxed microstructure is shown in \autoref{fig:vertex-micro}a. 

As the reference disorientation vectors follow a trivariate normal distribution, their norms follow a Maxwell distribution and in the limit of small angles so too does the distribution of reference disorientation angles $\omega$ (\autoref{fig:spread-ex}b). This kind of distribution provides a good first order approximation to orientation spreads found experimentally within deformed grains \cite{glez_orientation_2001,miodownik_mark_a._scaling_2001,pantleon_dislocation_2001}. 
In a similar way, boundary disorientations (i.e. disorientations calculated between pairs of spatially adjacent cells) are denoted $\delta\bm{r}^b=\left(r_1,r_2,r_3\right)^\mathit{b} sin\left(\theta/2\right)$.  As the cell orientations are spatially uncorrelated, the distribution of boundary disorientation angles also follows a Maxwell distribution\footnote{The development of a Maxwell distribution of disorientation angles during deformation of fcc metals has been shown to be a natural consequence of the operation of three orthogonal slip systems \cite{miodownik_mark_a._scaling_2001,pantleon_dislocation_2001}. Experimentally, the boundary disorientation angle distribution in fcc metals has
been found to be closer to a Rayleigh distribution \cite{hughes_scaling_1998}, implying the domination of slip by
two slip systems \cite{pantleon_dislocation_2001}} \cite{miodownik_mark_a._scaling_2001,pantleon_dislocation_2001,zecevic_modelling_2019}.




 Adopting a definition used in previous work \cite{holm_abnormal_2003,suwa_phase-field_2008}, recrystallized grains are defined as subgrains whose equivalent area radius is greater or equal to eight times the mean radius in the relaxed microstructure. \autoref{fig:vertex-micro}b shows the microstructure at 50\% recrystallized fraction. The exact value of the threshold recrystallized grain radius does not influence the comparison of the results as it will be set the same for the full-field and mean-field models.

\section{The mean-field model of cellular growth}


Following the approach of Humphreys \cite{humphreys_unified_1997} and Rollett and Mullins \cite{rollett_growth_1997}, the microstructure is considered in the mean-field model as a set of grains and subgrains embedded in a homogeneous medium representing the average properties of the microstructure. Growth rates of grains and subgrains are calculated from classic capillary growth laws, and a time-integration scheme is used to update the microstructure. At each time step, the mean boundary energies and mobilities required to compute growth rates are estimated from the moments of the boundary disorientation angle distribution. Here, the mean (first moment) and variance (centered second moment) of the disorientation angle distribution are estimated in a statistical sense from knowledge of the orientation spread and of potential spatial correlations between orientations. This approach differs from most traditional mean-field models, where boundaries properties are fixed from the start and recrystallized grains explicitly associated with a generic high angle boundary \cite{bailey_recrystallization_1962,zurob_quantitative_2006,huang_review_2016,zecevic_modelling_2019}. By considering the role of orientation spread on the determination of boundary properties, this model allows the recrystallization kinetics and recrystallized grain orientations to evolve together. 



%

%
\subsection{Growth equations and time integration}

We consider a set of individual cells characterized by radius $R_{\left(i,t\right)}$, mean boundary energy $\Gamma_{\left(i,t\right)}$ and mean boundary mobility $M_{\left(i,t\right)}$, embedded in a homogeneous medium of properties $\bar{R}_{\left(t\right)}$ and $\bar{\Gamma}_{\left(t\right)}$. The subscript $i$ denotes the cell index, and $t$ is the simulation time. Cells comprise all grains and subgrains in the microstructure, the same laws being applied to all objects. At $t=0$, the input radii correspond to the measurements in the as-relaxed full-field microstructure. On the other hand, the assignment of unique boundary properties to cells having multiple neighbours is one major approximation of this model, and will be treated below. Assuming that the above defined cell properties are known, the growth rate of a two dimensional cell is given by \cite{rollett_growth_1997}:

\begin{equation}
\frac{dR_{\left(i,t\right)}}{dt}=\frac{M_{\left(i,t\right)}\Gamma_{\left(i,t\right)}}{R_{\left(i,t\right)}}\left(\frac{a_{\left(i,t\right)}n_{\left(i,t\right)}}{6}-1\right)
\label{eq:growth-rate}
\end{equation}

Where $n_{\left(i,t\right)}$ is the number of sides (or neighbours) of the cell, and $a_{\left(i,t\right)}=6sin^{-1}\left(\bar{\Gamma}_{\left(t\right)}/2\Gamma_{\left(i,t\right)}\right)/\pi\leq 3$ accounts for the effect of boundary curvature on the growth rate. For two-dimensional microstructures, a linear relation can be assumed between the number of sides of a cell and its size such that $n_{\left(i,t\right)}=3\left(1+R_{\left(i,t\right)}/\bar{R}_{\left(t\right)}\right)$ \cite{rollett_growth_1997,hillert_theory_1965}. The growth rate equation is thus reduced to\footnote{We provide in \autoref{app:3DMullins} a similar expression of growth rates for microstructures in 3 dimensions.} \cite{rollett_growth_1997}:
\begin{equation}
\frac{dR_{\left(i,t\right)}}{dt}=\frac{M_{\left(i,t\right)}\Gamma_{\left(i,t\right)}}{2R_{\left(i,t\right)}}\left(a_{\left(i,t\right)}\left(1+\frac{R_{\left(i,t\right)}}{\bar{R}_{\left(t\right)}}\right)-2\right)
\label{eq:growth-rate2}
\end{equation}


Then each cell radius is updated by integrating \autoref{eq:growth-rate2} with Euler's method $R_{\left(i,t+dt\right)}=R_{\left(i,t\right)}+\frac{dR_{\left(i,t\right)}}{dt}\Delta t$. The model's predictions were found insensitive to the choice of $\Delta t$ so long as the average increase in cell area per time increment remained below $\sim$1\%. Then, recrystallized grains are identified, as in the full-field simulations, based on a critical radius $R_{\left(i,t\right)}\geq R_\mathit{rx}=8\bar{R}_{\left(0\right)}$. 

After each time increment, the smallest cells and those of negative radius are removed in order to maintain a constant total simulation area. This procedure is implemented to compensate for the fact that \autoref{eq:growth-rate2} (or \autoref{eq:growth-rate}) does not intrinsically insure area conservation (i.e. $\sum_{i=1}^{n_g\left(t\right)}R_{\left(i,t+dt\right)}\frac{dR_{\left(i,t\right)}}{dt}\ne 0$). To correct this discrepancy, one may further refine the contribution of the homogeneous medium to the growth rates \cite{abbruzzese_theory_1986}, but this does not change the results presented below. In all cases, the total change in area was not more than 8\%.

%

\subsection{Boundary properties}

To compute the growth rates in \autoref{eq:growth-rate2}, one needs to know the mean boundary energy terms $\Gamma_{\left(i,t\right)}$ and $\bar{\Gamma}_{\left(i,t\right)}$ and mean mobility $M_{\left(i,t\right)}$. Here, we estimate them in two steps. First the moments of the boundary disorientation angle distribution are estimated statistically from those associated with the reference disorientation distribution (i.e. the orientation spread) and from assumptions on the spatial correlations. Then, boundary properties are calculated assuming Taylor series expansion of the energy and mobility laws about the mean boundary disorientation angles. 

First, the moments of the reference disorientation distribution are calculated in a statistical sense. As such, they carry no information on the neighbour to neighbour disorientations. The first moment is $\left(0,0,0\right)$ since orientations are centered on the average. The second moment is a 3x3 matrix given by \cite{pantleon_retrieving_2005}:


\begin{equation}
<\delta\bm{r}^\mathit{ref}\otimes\delta\bm{r}^\mathit{ref}>_{\left(t\right)}=\frac{1}{n_g\left(t\right)}
\sum_{i=1}^{n_g\left(t\right)}\delta\bm{r}_{\left(i\right)}^\mathit{ref}\otimes\delta\bm{r}_{\left(i\right)}^\mathit{ref}
\label{eq:discrete-smom}
\end{equation}
Where $<>_{\left(t\right)}$ denotes the average of the quantity within brackets at time $t$, $\otimes$ is the dyadic product, and the sum runs over the $n_g\left(t\right)$ orientations\footnote{Weighting the second moment based on size, i.e. $\frac{1}{\sum_{i=1}^{n_g\left(t\right)}R_{\left(i\right)}}\sum_{i=1}^{n_g\left(t\right)}R_{\left(i\right)}\delta\bm{r}_i^\mathit{ref}\otimes\delta\bm{r}_i^\mathit{ref}$ did not lead to improved predictions.}. Since the first moment is null, the second moment is also the covariance matrix of the distribution. Its eigenvalues provide the square of the standard deviations $\left(\sigma^1_{\left(t\right)}, \,\sigma^2_{\left(t\right)},\,\sigma^3_{\left(t\right)}\right)$ of spread in the principal directions of the reference frame. With an isotropic spread $\sigma^\mathit{ref}_{\left(t\right)}=\sigma^1_{\left(t\right)}=\sigma^2_{\left(t\right)}=\sigma^3_{\left(t\right)}$. 



Next, one can estimate the moments of the boundary disorientation vector distribution from those associated with reference disorientations. Zecevic \emph{et al.} \cite{zecevic_modelling_2019} have shown that when the reference disorientation vectors follow a trivariate normal distribution, the second moment of the boundary disorientation vector distribution for a cell of reference disorientation $\delta\bm{r}_{\left(i\right)}^\mathit{ref}$ can be expressed by:

\begin{equation}
<\delta\bm{r}^\mathit{b}\otimes\delta\bm{r}^\mathit{b}>_{\left(i,t\right)}=\delta\bm{r}_{\left(i\right)}^\mathit{ref}\otimes\delta\bm{r}_{\left(i\right)}^\mathit{ref}+\left(<\delta\bm{r}^\mathit{ref}\otimes\delta\bm{r}^\mathit{ref}>_{\left(t\right)}^{\,-1}+\frac{1}{\alpha}\bm{I}\right)^{-1}
\label{eq:rawsmom-indiv}
\end{equation}

Where $\bm{I}$ is a 3$\times$3 identity matrix, and $\alpha$ is the variance of a spatial correlation function of Gaussian form. This parameter ranges from 0 for high spatial correlation (i.e. when grains and subgrains of similar orientation are most likely to be adjacent) to $+\infty$ for no correlation. For our case, where we assume no spatial correlations between orientations, $\alpha\rightarrow+\infty$, and \autoref{eq:rawsmom-indiv} simplifies to:

\begin{equation}
<\delta\bm{r}^\mathit{b}\otimes\delta\bm{r}^\mathit{b}>_{\left(i,t\right)}=\delta\bm{r}_{\left(i\right)}^\mathit{ref}\otimes\delta\bm{r}_{\left(i\right)}^\mathit{ref}+<\delta\bm{r}^\mathit{ref}\otimes\delta\bm{r}^\mathit{ref}>_{\left(t\right)}
\label{eq:rawsmom-indiv2}
\end{equation}

In \autoref{eq:rawsmom-indiv} and \autoref{eq:rawsmom-indiv2}, the first term on the right represents the shift of the cell orientation from the reference frame, while the second term is the covariance matrix of the boundary disorientation distribution. In the case of \autoref{eq:rawsmom-indiv2}, the covariance matrices of the reference and boundary disorientation vectors are identical, and the isotropic spread of the boundary disorientation distribution becomes $\sigma^\mathit{b}_{\left(i,t\right)}=\sigma^\mathit{ref}_{\left(t\right)}$. 

As the boundary disorientation vectors follow a trivariate normal distribution with non-zero mean, the variable $\Theta_{\left(i,t\right)}=\sqrt{\delta\bm{r}^\mathit{b}_{\left(i,t\right)}\cdot\delta\bm{r}^\mathit{b}_{\left(i,t\right)}}/\sigma^b_{\left(i,t\right)}$ follows a non-central $\chi$ distribution \cite{park_moments_1961}. In addition, in the hypothesis of small angles, $\theta_{\left(i,t\right)}\approx2\sqrt{\left(\delta\bm{r}^\mathit{b}_{\left(i,t\right)}\cdot\delta\bm{r}^\mathit{b}_{\left(i,t\right)}\right)}$. With $\theta_{\left(i,t\right)}$ and $\Theta_{\left(i,t\right)}$ being proportional, the moments of the boundary disorientation angle distributions can be expressed as a function of the moments of the $\chi$ distribution for the variable $\Theta_{\left(i,t\right)}$:



\begin{equation}
<\theta>_{\left(i,t\right)}=2\sigma^\mathit{b}_{\left(i,t\right)}\sqrt{\frac{\pi}{2}}L_{1/2}^{\left(1/2\right)}\left(-\frac{\left(\kappa_{\left(i,t\right)}\right)^2}{2}\right)
\label{eq:Eindiv}
\end{equation}

\begin{equation}
<\theta^2>_{\left(i,t\right)}=\left(2\sigma^\mathit{b}_{\left(i,t\right)}\right)^2\left(3+\left(\kappa_{\left(i,t\right)}\right)^2\right)
\label{eq:Vindiv}
\end{equation}

Where $L_{1/2}^{\left(1/2\right)}$ is the Laguerre function of coefficients ${1/2}$ and $\left(1/2\right)$ (see \autoref{app:Laguerre}), and $ \kappa_{\left(i,t\right)}$ is a scaling parameter defined by:

\begin{equation}
    \kappa_{\left(i,t\right)}=\frac{\sqrt{\delta\bm{r}_{\left(i\right)}^\mathit{ref}\cdot\delta\bm{r}_{\left(i\right)}^\mathit{ref}}}{\sigma^\mathit{b}_{\left(i,t\right)}}\approx\frac{{\omega}}{2\sigma^\mathit{b}_{\left(i,t\right)}}
    \label{eq:kappa-indiv}
\end{equation}

The approximation on the right side of the equation holds for small angles, and is shown only to highlight the dependency on the disorientation angle $\omega$. Since the isotropic spread $\sigma^\mathit{b}_{\left(i,t\right)}$ is the same for all cells and equal to $\sigma^\mathit{ref}_{\left(t\right)}$, the parameter $\kappa_{\left(i,t\right)}$ is constant for a given reference disorientation angle $\omega$. 

%



The same method can be used to estimate the moments of the boundary disorientation vector distribution taken over the whole microstructure. This distribution is centered on $\left(0,0,0\right)$, and its second moment is \cite{zecevic_modelling_2019}:

\begin{multline}
<\delta\bm{r}^\mathit{b}\otimes\delta\bm{r}^\mathit{b}>_{\left(\forall i,t\right)}=\left(<\delta\bm{r}^\mathit{ref}\otimes\delta\bm{r}^\mathit{ref}>_{\left(t\right)}^{\,-1}+\frac{1}{\alpha}\bm{I}\right)^{-1}<\delta\bm{r}^\mathit{ref}\otimes\delta\bm{r}^\mathit{ref}>_{\left(t\right)}^{\,-T}
\\
\left(<\delta\bm{r}^\mathit{ref}\otimes\delta\bm{r}^\mathit{ref}>_{\left(t\right)}^{\,-1}+\frac{1}{\alpha}\bm{I}\right)^{-T}+\left(<\delta\bm{r}^\mathit{ref}\otimes\delta\bm{r}^\mathit{ref}>_{\left(t\right)}^{\,-1}+\frac{1}{\alpha}\bm{I}\right)^{-1}
\label{eq:rawsmom-all}
\end{multline}

Where $\left(\forall i\right)$ indicates that the average is now taken for all cells in the microstructure, and $-T$ is the inverse of the transpose matrix. Considering, again, that $\alpha\rightarrow+\infty$, \autoref{eq:rawsmom-all} simplifies to:

\begin{equation}
<\delta\bm{r}^\mathit{b}\otimes\delta\bm{r}^\mathit{b}>_{\left(\forall i,t\right)}=2<\delta\bm{r}^\mathit{ref}\otimes\delta\bm{r}^\mathit{ref}>_{\left(t\right)}
\label{eq:rawsmom-all2}
\end{equation}

The isotropic spread of boundary disorientation vectors for the whole microstructure is thus equal to $\sigma^b_{\left(\forall i,t\right)}=\sqrt{2}\sigma^\mathit{ref}_{\left(t\right)}$. Since the distribution is centered on $\left(0,0,0\right)$, $ \kappa_{\left(\forall i,t\right)}=0$, and the first and second moments of boundary disorientation angles follow \footnote{Note than when $\kappa=0$, the $\chi$ distribution coincides with the Maxwell distribution cited earlier.}:

\begin{equation}
<\theta>_{\left(\forall i,t\right)}=2\sigma^b_{\left(\forall i,t\right)}\sqrt{\frac{\pi}{2}}L_{1/2}^{\left(1/2\right)}\left(0\right)
\label{eq:meandiso}
\end{equation}

\begin{equation}
<\theta^2>_{\left(\forall i,t\right)}=12\left(\sigma^b_{\left(\forall i,t\right)}\right)^2
\end{equation}

Finally, the mean boundary energies and mobilities can be estimated from the moments of the boundary disorientation angle distributions. Expressing the mobility and energy laws by a second order Taylor series about the mean boundary disorientations gives the mean boundary mobilities and energies:

\begin{equation}
M_{\left(i,t\right)}=<\mu\left(\theta\right)>_{\left(i,t\right)}=\mu\left(<\theta>_{\left(i,t\right)}\right)+\frac{\mu''\left(<\theta>_{\left(i,t\right)}\right)}{2}\left(<\theta^2>_{\left(i,t\right)}-\left(<\theta>_{\left(i,t\right)}\right)^2\right)
\label{eq:M}
\end{equation}

\begin{equation}
\Gamma_{\left(i,t\right)}=<\gamma\left(\theta\right)>_{\left(i,t\right)}=\gamma\left(<\theta>_{\left(i,t\right)}\right)+\frac{\gamma''\left(<\theta>_{\left(i,t\right)}\right)}{2}\left(<\theta^2>_{\left(i,t\right)}-\left(<\theta>_{\left(i,t\right)}\right)^2\right)
\label{eq:Gamma}
\end{equation}

\begin{equation}
\bar{\Gamma}_{\left(t\right)}=<\gamma\left(\theta\right)>_{\left(\forall i,t\right)}=\gamma\left(<\theta>_{\left(\forall i,t\right)}\right)+\frac{\gamma''\left(<\theta>_{\left(\forall i,t\right)}\right)}{2}\left(<\theta^2>_{\left(\forall i,t\right)}-\left(<\theta>_{\left(\forall i,t\right)}\right)^2\right)
\label{eq:Gamma-bar}
\end{equation}


As will be shown below, the second order terms are not necessary to capture the main trends of the microstructure evolution, but they substantially increase the accuracy of the prediction. The second derivatives of the mobility and energy functions are provided in \autoref{app:2derivative}. 

\subsection{Algorithm}

First, the model reads as input a list of subgrains characterized by their radii and orientations. Since an orientation is characterized by at least three parameters regardless of the representation space, the total is four parameters per subgrain. From the list of subgrain orientations, the reference orientation
is computed as well as the reference disorientations. The initial simulation area is obtained from the sum of all subgrain areas in the input microstructure. In this case study, the input files were generated from the cell parameters of the relaxed Vertex microstructures. Other potential sources of input will be discussed below.

Next, the time iteration loop is started. From time $t$ to $t+dt$, the following sequence is executed:

\begin{enumerate}
    \item Identify the recrystallized grains based on the size criterion.
    \item Calculate the mean boundary properties $M_{\left(i,t\right)}$ and $\Gamma_{\left(i,t\right)}$ for each cell, and $\bar{\Gamma}_{\left(t\right)}$ for the whole microstructure using \autoref{eq:M}, \autoref{eq:Gamma} and \autoref{eq:Gamma-bar}.
    \item Calculate the growth rate of each cell using \autoref{eq:growth-rate2}.
    \item Integrate the growth rates over a time increment to update the cell radii. The new cell radii are representative of the microstructure at time $t+dt$.
    \item Remove the cells of negative radius and the smallest cells of positive radius so as to minimize the difference between the initial microstructure area and the sum of cell areas at $t+dt$.
    \item Update the average radius.
\end{enumerate}

The only model parameter is the variance of the spatial correlation function $\alpha$, set to $+\infty$ (no correlation) in this case. The parameters controlling the boundary energy and mobility laws were set identical to the full-field simulation. Any other mobility and energy laws can be implemented as long as they are differentiable to the second order. This implementation is called the complete mean-field model for the rest of the paper.


\section{Results}

In this section, the mean-field model predictions are compared to a full-field simulation of recrystallization realized with an initial orientation spread of $\sigma^{\textit{ref}}_{\left(0\right)}$=3.5$^\circ$. This value is in the range of experimental measurements in deformed polycrystalline materials \cite{krog-pedersen_quantitative_2009,despres_contribution_2020}. The initial subgrain number density is denoted $\rho_0$. This parameter is used as a normalizing factor in much of the subsequent analysis.

To highlight the role of the different components of the mean-field model to the prediction of recrystallization parameters, four different ways of calculating the boundary properties are compared: 

\begin{enumerate}
\item Using only mean boundary disorientation angles (i.e. 0$^\text{th}$ order Taylor series expansion), kept fixed with time and calculated initially at $t=0$.
\item Using time-updated mean boundary disorientation angles.
\item Using means and variances of the boundary disorientation angles (i.e. 2$^\text{nd}$ order Taylor series expansion), kept fixed with time and calculated initially at $t=0$.
\item Using time-updated means and variances of the boundary disorientation angles (i.e. complete model).
\end{enumerate}

\subsection{Prediction of recrystallization kinetics}

\begin{figure}[htbp]
\centering

{\includegraphics[width=0.49\textwidth]{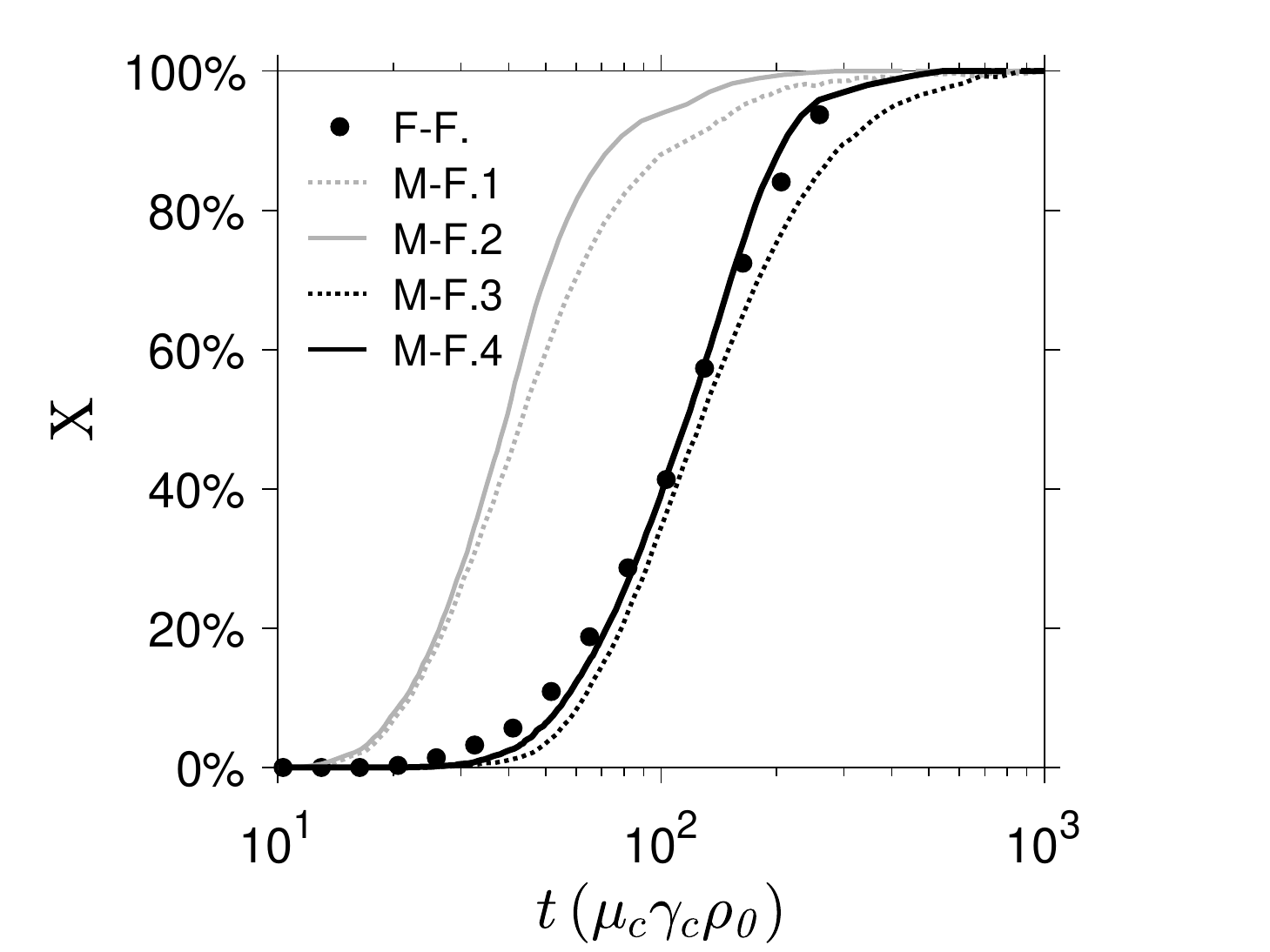}}

\caption{Comparison of the recrystallization kinetics predicted by the mean- and full-field models. Time is normalized by $1/\left(\mu_c\gamma_c\rho_{0}\right)$. Full-field simulations performed with the Vertex model  appear as points (F-F.). Grey lines denote mean-field predictions made using only mean disorientation angles, either fixed (M-F.1) or time-updated (M-F.2). Black lines denote mean-field predictions made using means and variances of the disorientation angles, either fixed (M-F.3) or time-updated (M-F.4, i.e. complete model).}
\label{fig:kinetics}
\end{figure}

To illustrate the influence of boundary properties and their time-evolution on the microstructure, \autoref{fig:kinetics} compares the full-field simulation of recrystallization kinetics to the four variants of the mean-field simulations. The recrystallized fraction $X$ is defined as the area of recrystallized grains divided by the total simulation area. The dotted grey line shows the predicted kinetics when the boundary properties are calculated using mean disorientation angles and are not updated with time. As a result, the recrystallization kinetics are overpredicted compared to the full-field simulation. Hurley and Humphreys arrived at the same result with similar assumptions in their mean-field model \cite{hurley_application_2003}. The solid grey line shows results obtained when the boundary properties are calculated using mean disorientation angles, but under the conditions that they are updated with time. The mean-field model predictions are not significantly improved.


\begin{figure}[htbp]
\centering
\subfigure[]
{\includegraphics[width=0.49\textwidth]{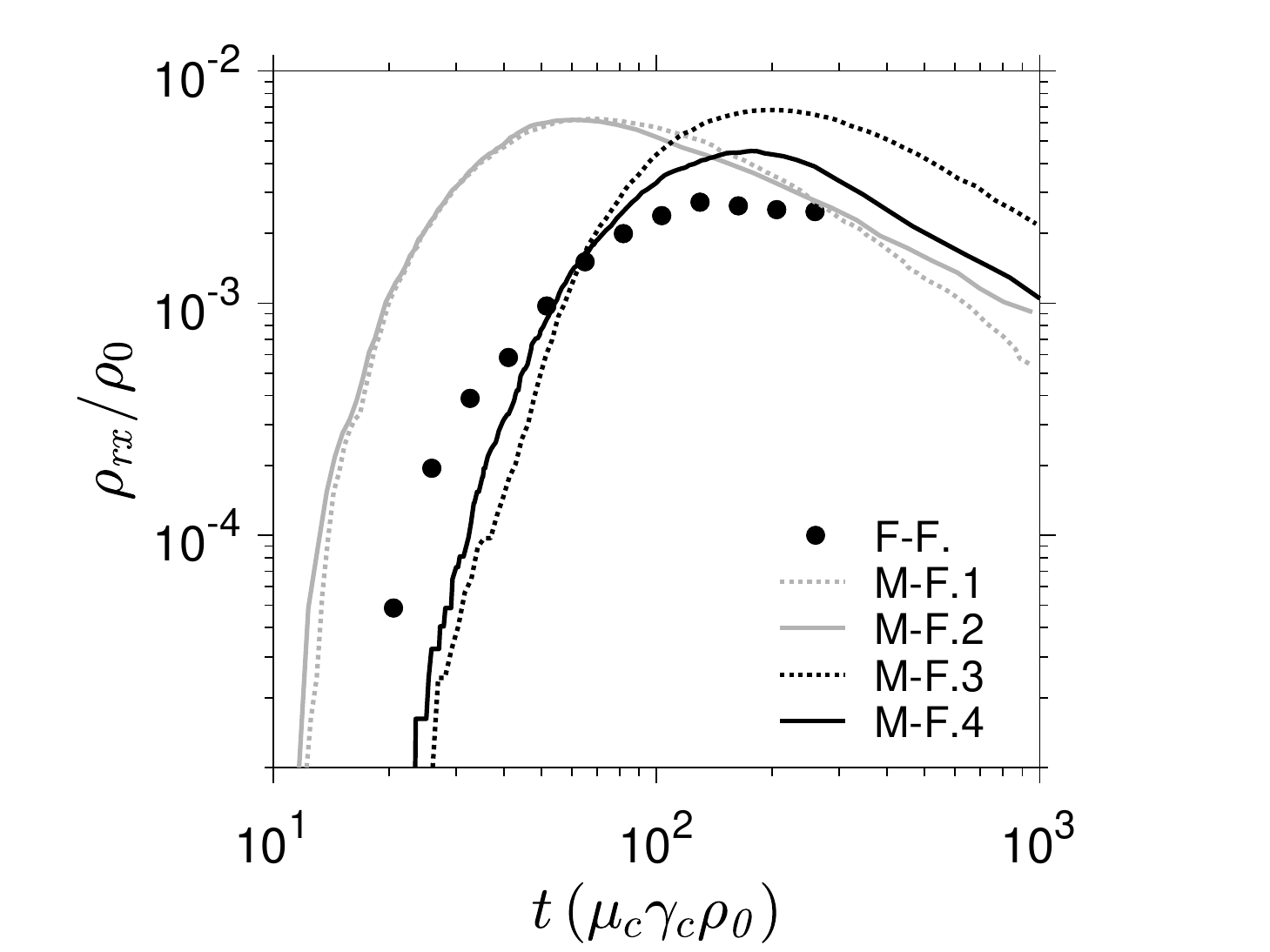}}
\subfigure[]
{\includegraphics[width=0.49\textwidth]{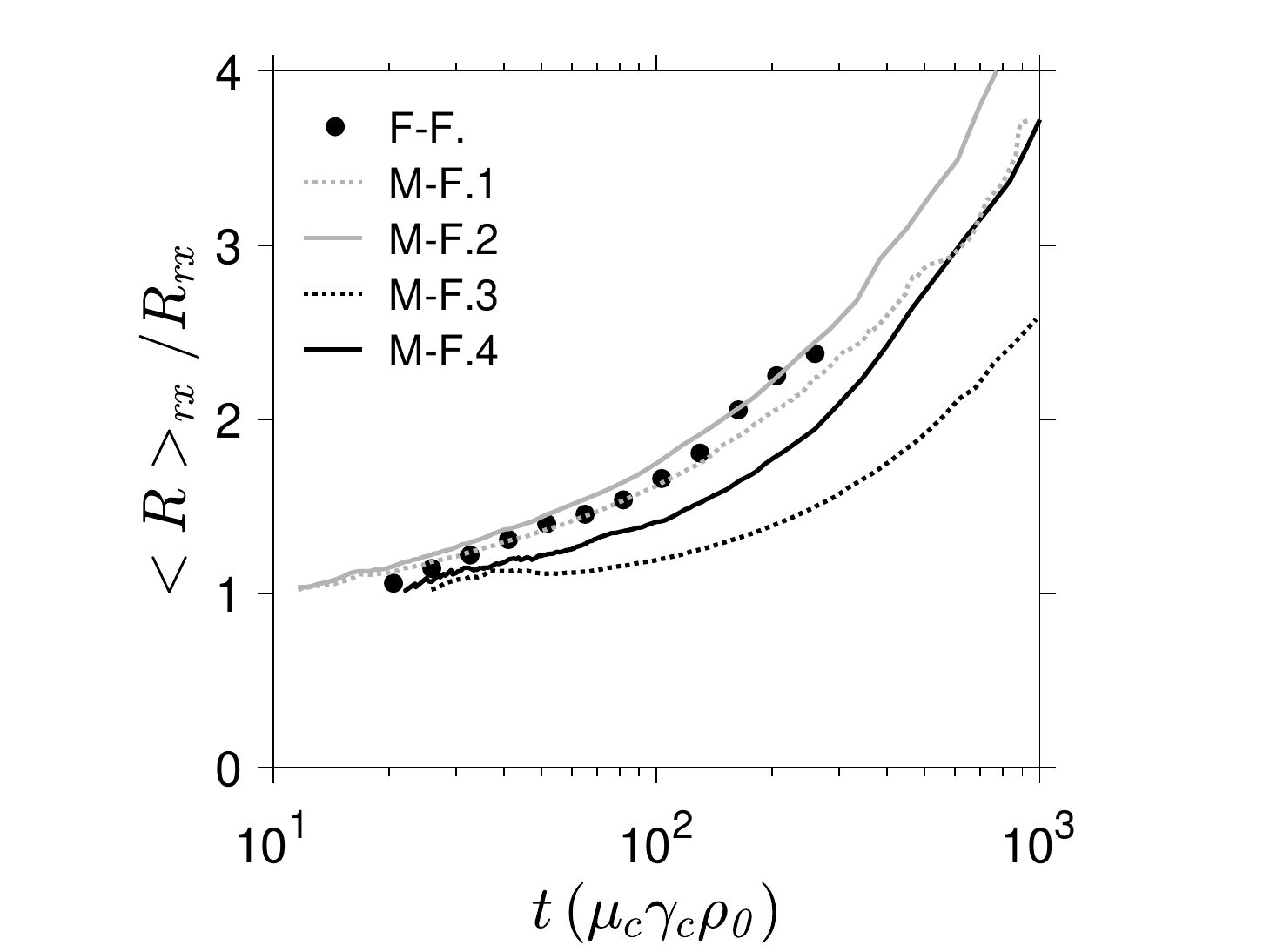}}

\caption{Comparison of a) the recrystallized grain density $\rho_\mathit{rx}$ and b) the mean recrystallized grain radius $<R>_\mathit{rx}$ predicted by the mean- and full-field models. Time is normalized by $1/\left(\mu_c\gamma_c\rho_{0}\right)$, recrystallized grain density by $\rho_0$ and recrystallized grain radius by the threshold recrystallized grain radius $R_\mathit{rx}$.  Full-field simulations performed with the Vertex model appear as points (F-F.). Grey lines denote mean-field predictions made using only mean disorientation angles, either fixed (M-F.1) or time-updated (M-F.2). Black lines denote mean-field predictions made using means and variances of the disorientation angles, either fixed (M-F.3) or time-updated (M-F.4, i.e. complete model).}
\label{fig:rhorx}
\end{figure}

A much better agreement is found between full-field and mean-field simulations when boundary energies and mobilities are calculated using both the means and variances of the boundary disorientation angle distributions (solid and dotted black lines). Updating the boundary properties is beneficial but of second order for the prediction of kinetics. In this case, the time at 50\% recrystallized fraction is predicted with less than 2\% error between the full-field simulation and the mean-field model prediction. Neither expanding the series expansion beyond the 2$^\text{nd}$ order in \autoref{eq:M}, \autoref{eq:Gamma} and \autoref{eq:Gamma-bar}, nor performing the series expansion directly on growth rates significantly improved the predictions.  


The same analysis is performed in  \autoref{fig:rhorx} for the recrystallized grain density and size. These parameters are of interest as they directly determine the recrystallization kinetics. \autoref{fig:rhorx}a shows that the best agreement with the full-field simulation in terms of recrystallized grain density $\rho_\mathit{rx}$ is obtained with the complete mean-field model (solid black line). The differences between the full-field simulation and the four mean-field model predictions mirror those of the recrystallization kinetics shown in \autoref{fig:kinetics}. In particular, one can see that only using the mean boundary disorientation angles to predict boundary properties (grey lines) significantly overpredicts the number of recrystallized grains and the time required for their appearance. This helps to explain the overpredicted kinetics in \autoref{fig:kinetics}. The decrease in recrystallized grain density predicted by all implementations at long times corresponds to the impingement of recrystallized grains at the end of recrystallizaton. This non-monotonic evolution has been observed experimentally in aluminium alloys by Perryman \emph{et al.} \cite{perryman_recrystallization_1955}.

 \autoref{fig:rhorx}b shows that the complete mean-field model (solid black line) predicts correctly the the mean recrystallized grain radius $<R>_\mathit{rx}$. One may notice that the implementations relying only on mean boundary disorientation angles (grey lines) yield even better predictions. This results from the faster predicted recrystallization kinetics, which translates the curve of recrystallized grain radius towards short times.

\subsection{Prediction of recrystallized grain orientations}

\begin{figure}[htbp]
\centering
\subfigure[]
{\includegraphics[width=0.49\textwidth]{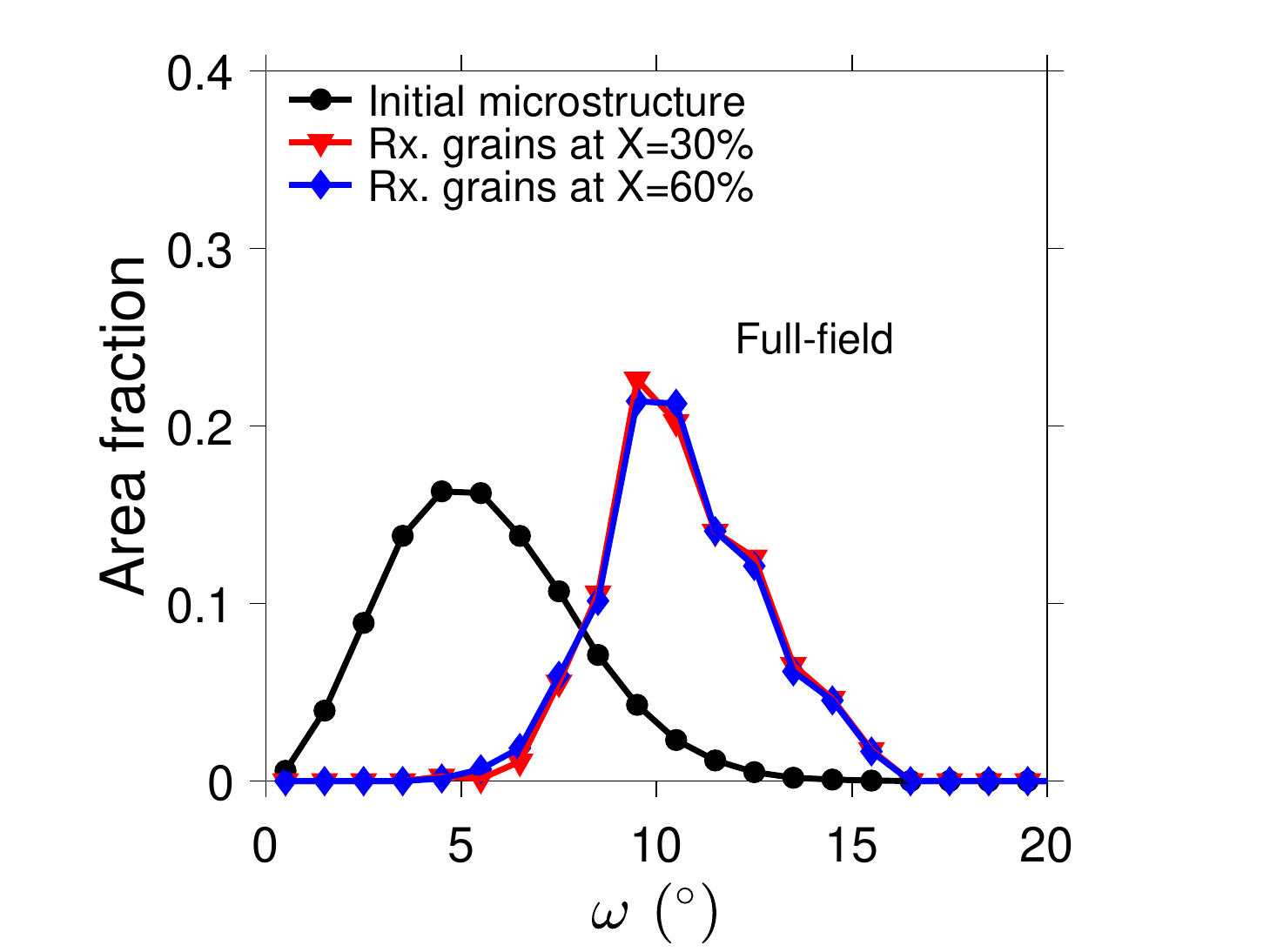}}

\subfigure[]
{\includegraphics[width=0.49\textwidth]{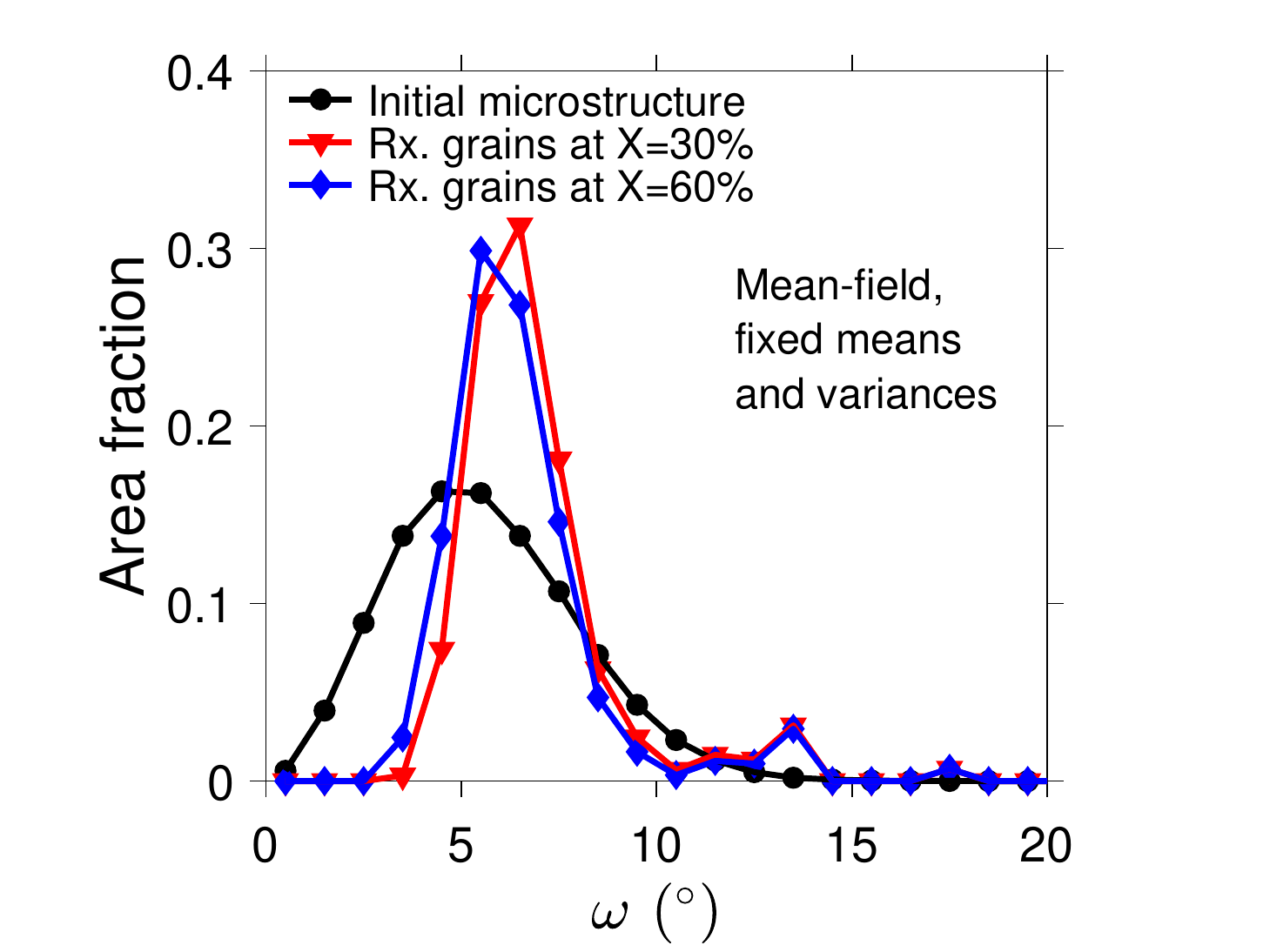}}
\subfigure[]
{\includegraphics[width=0.49\textwidth]{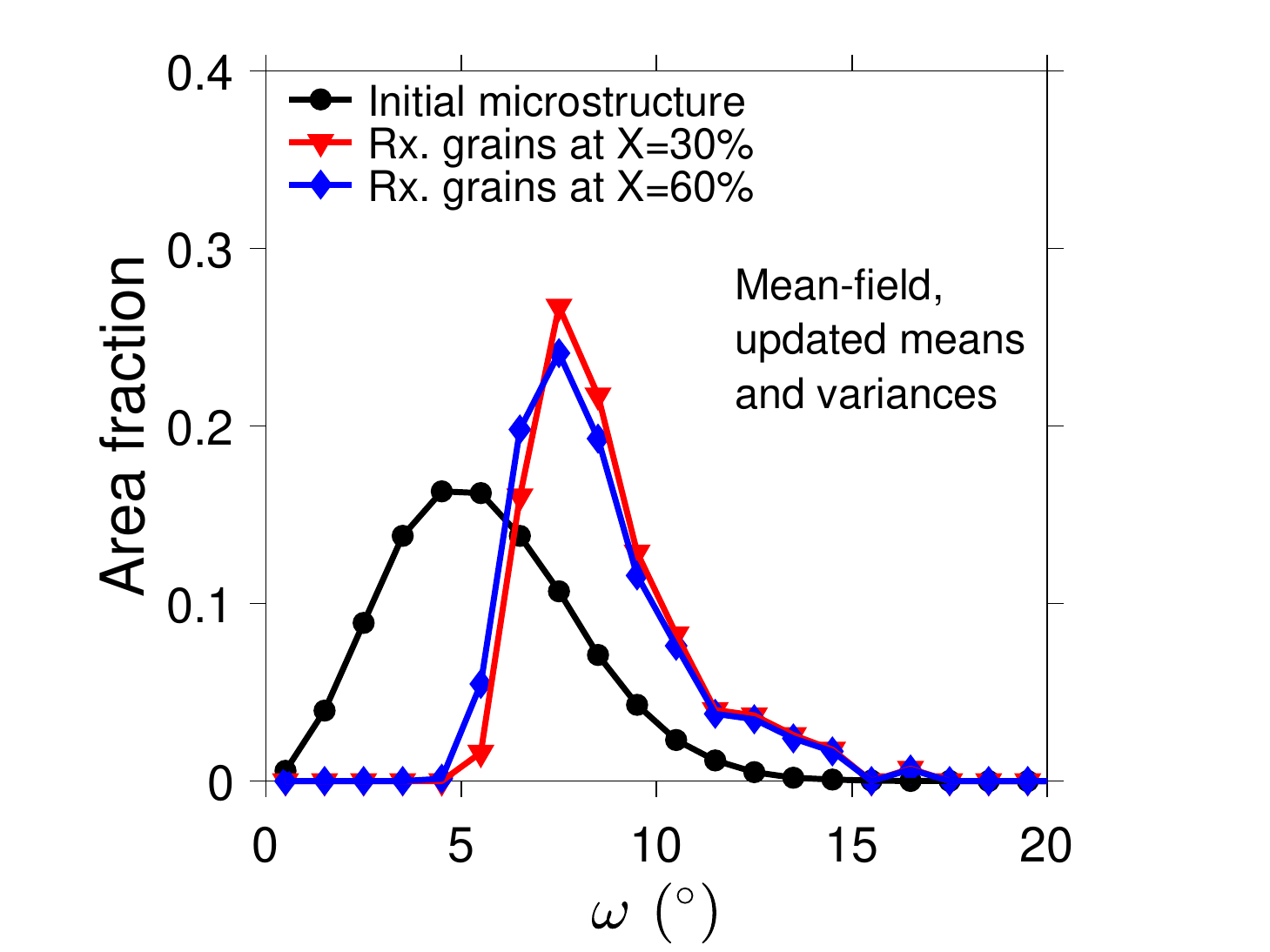}}
\caption{Area fraction of grains and subgrains as a function of their reference disorientation
angle $\omega$. a) Full-field simulation, b) mean-field simulation considering means and variances of boundary disorientation angle distribution fixed at their initial values, c) mean-field model considering means and variances of boundary disorientation angle distribution updated with time. The initial microstructure dataset (in black) includes all grains and subgrains regardless of recrystallization, and is identical for a), b) and c). The recrystallized grain datasets (red and blue) include only the recrystallized grains at specific recrystallized fractions. }
\label{fig:PF}
\end{figure}

While the time evolution of boundary properties influences only moderately the recrystallization kinetics, it is a critical aspect for determining the recrystallized grain orientations. To illustrate this, \autoref{fig:PF} compares the distribution of grain and subgrain orientations predicted by the full-field model and by the two implementations of the mean field model using means and variances of boundary disorientation angle distributions (i.e. 2$^\text{nd}$ Taylor series expansion with fixed and updated boundary properties). As the orientation spread is isotropic, orientations
can be plotted as a function of their reference disorientation angle $\omega$. For the full-field simulations (\autoref{fig:PF}a), a preferential development of orientations
with large reference disorientation angle (i.e. with approximately $\omega\geq5^\circ$) in the recrystallized grains is observed. The orientations that develop with highest fraction exhibit a compromise between fraction in the initial microstructure and magnitude of the disorientation angle. Without updating the boundary properties with time, the mean-field model poorly captures this evolution (\autoref{fig:PF}b). The agreement is improved when updating the properties with time (\autoref{fig:PF}c). While in this case the mode of the distribution for the recrystallized grains is still lower than that simulated by the full-field model, the range of disorientation angles is very well captured.


In summary, the complete mean-field model yields the best predictions of recrystallization kinetics and grain orientations while also giving a good representation of the evolution of the mean recrystallized grain size. This implementation is kept for all following analyses.

\subsection{Prediction of boundary properties}



\begin{figure}[htbp]
\centering
\subfigure[]{\includegraphics[width=0.49\textwidth]{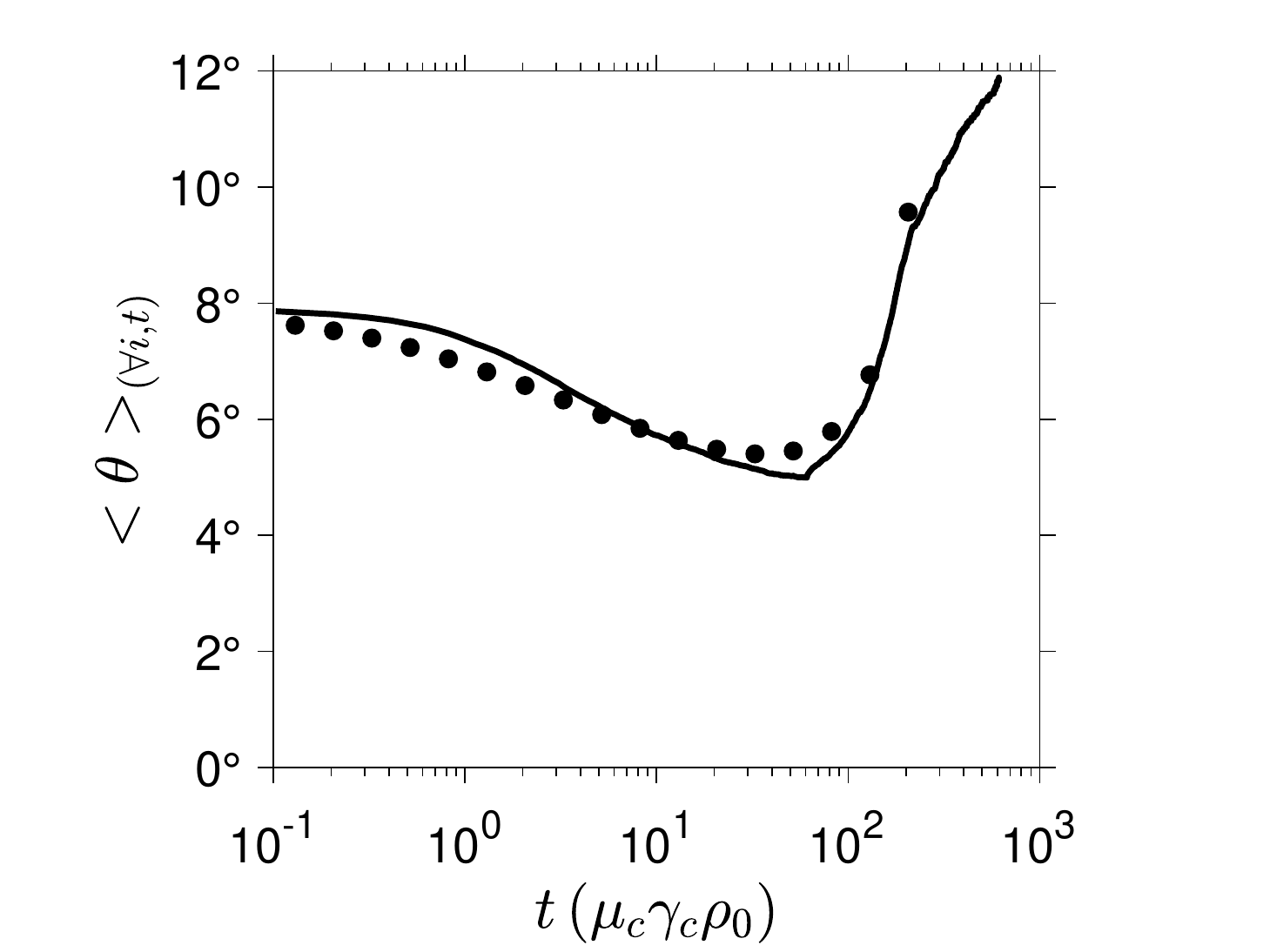}}
\subfigure[]{\includegraphics[width=0.49\textwidth]{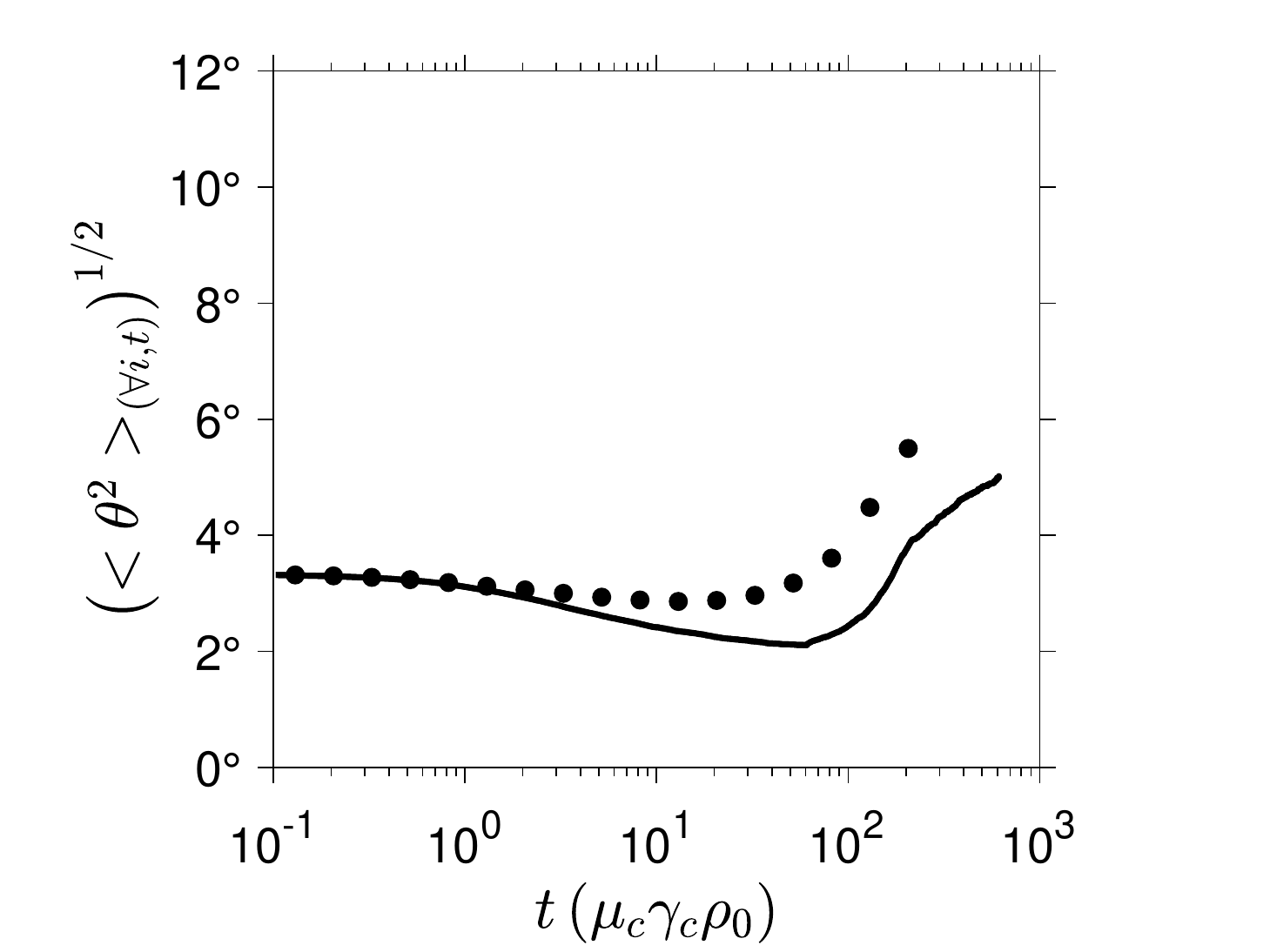}}
\caption{a) first moment and b) square root of the second moment of the boundary disorientation distribution as a function of time. Time is normalized by $1/\left(\mu_c\gamma_c\rho_{0}\right)$. Points are calculated from the list of boundary properties in the full-field simulation, while lines are the mean-field model predictions.}
\label{fig:global-spread}
\end{figure}

To evaluate the success of the mean-field model in predicting boundary properties, one can directly track them in the full-field and complete mean-field models. \autoref{fig:global-spread} shows that the evolution of the first and second moments of the boundary disorientation angle distribution is well captured by the mean-field model. Both moments evolve in a non-monotonic way, comparable to experimental observations\footnote{Huang and Humphreys \cite{huang_subgrain_2000} reported a decrease of the mean boundary disorientation during annealing of a deformed aluminium monocrystal. Mishin \emph{et al.} \cite{mishin_recovery_2013} showed a similar decrease then increase of the density of high angle boundaries during static recrystallization of a polycristalline aluminium alloy.}. The magnitudes are well captured by the mean-field model, although the second moment tends to be underpredicted at longer times. 





\begin{figure}[htbp]
\centering
\subfigure[]{\includegraphics[width=0.49\textwidth]{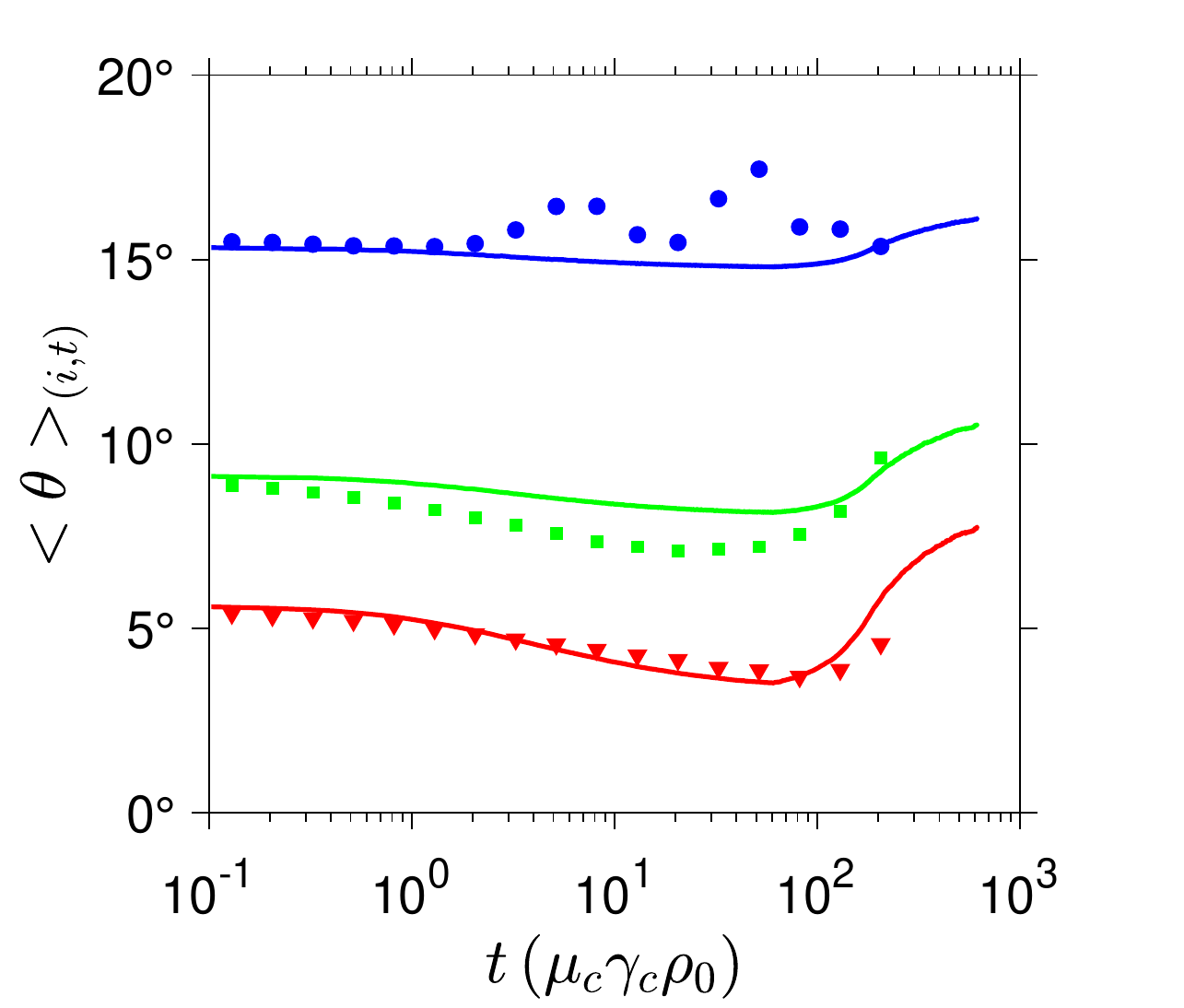}}
\subfigure[]{\includegraphics[width=0.49\textwidth]{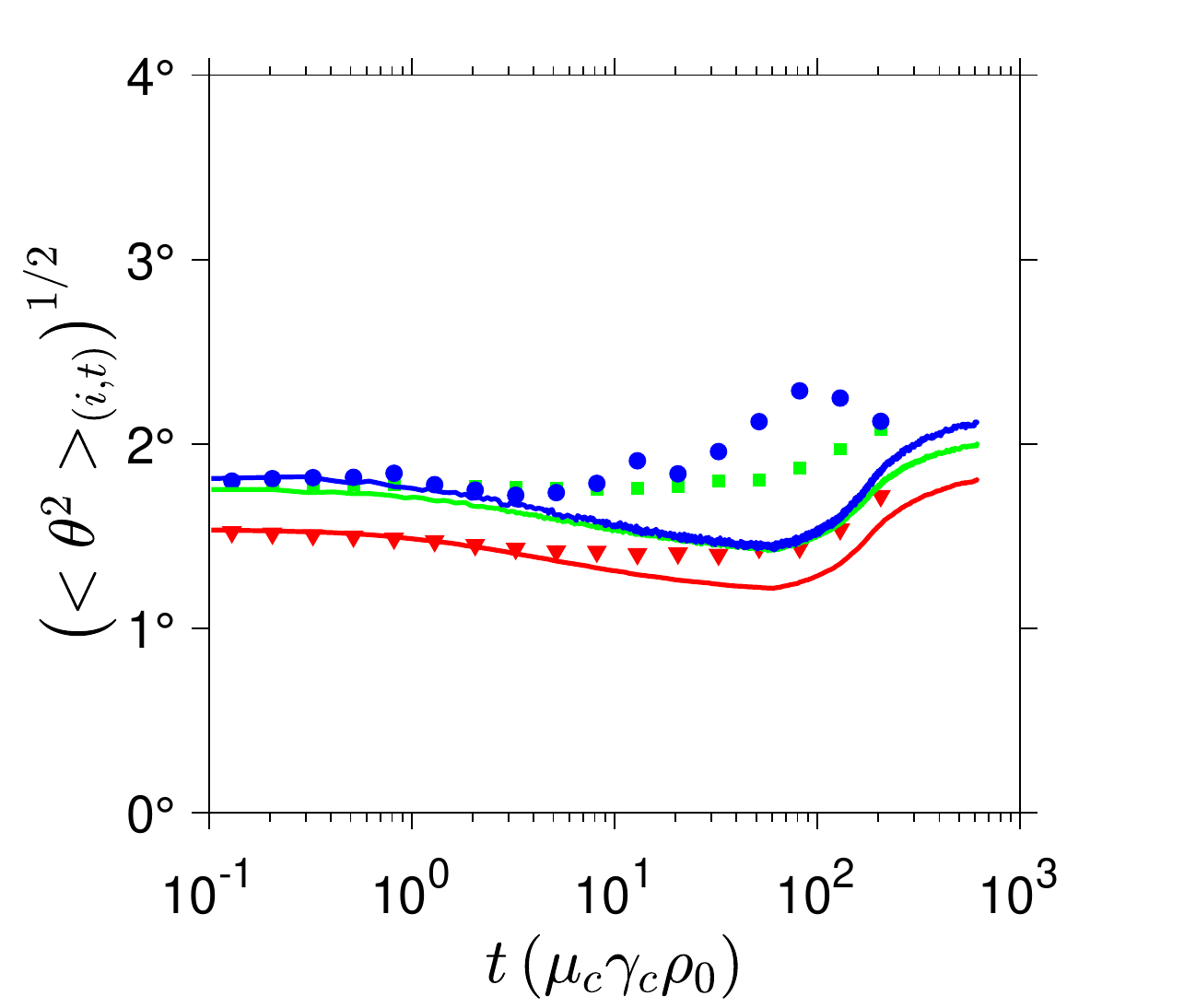}}

\subfigure[]{\includegraphics[width=0.49\textwidth]{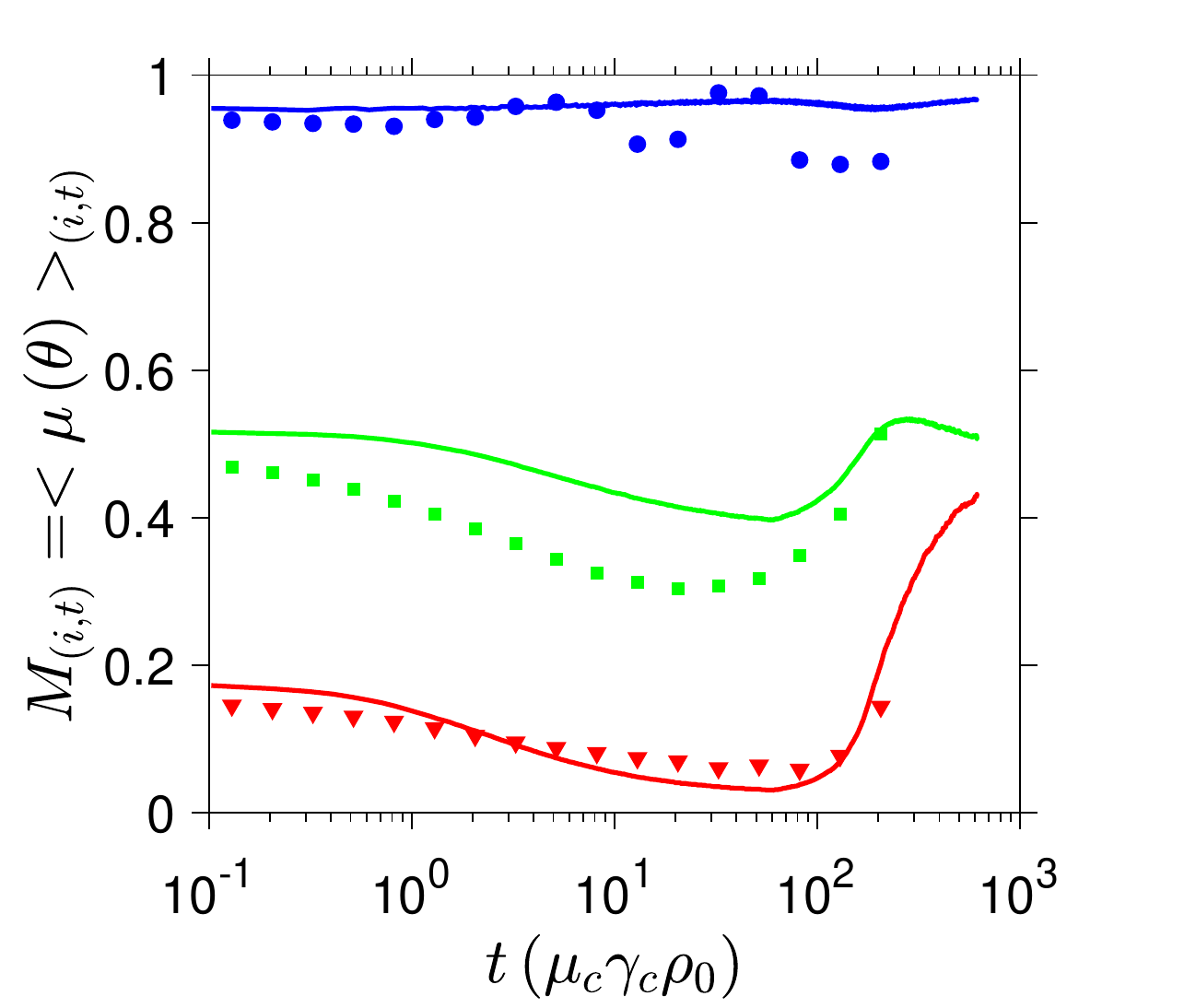}}
\subfigure[]{\includegraphics[width=0.49\textwidth]{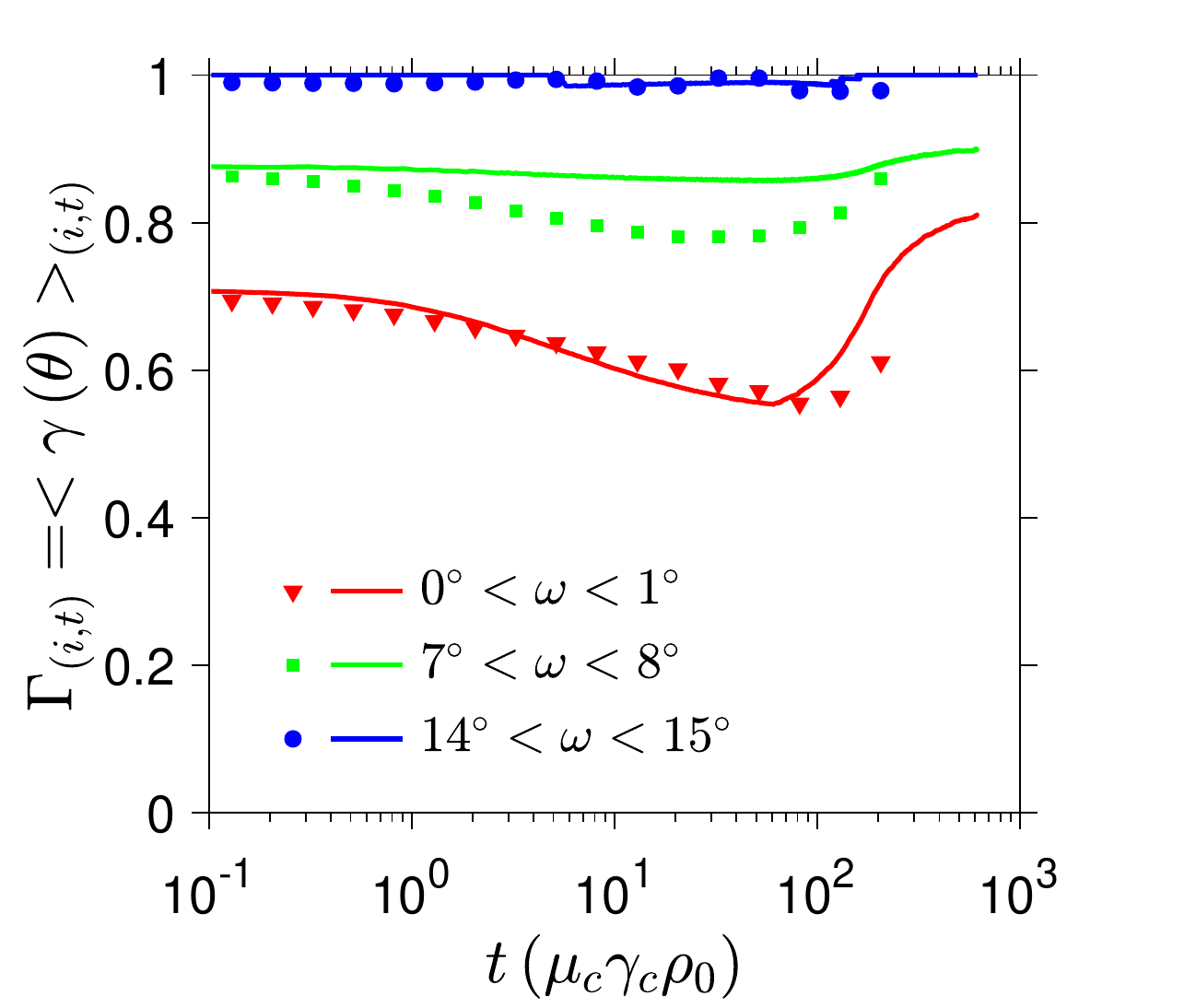}}

\caption{Cell boundary properties as a function of time for three intervals of reference disorientation angle $\omega$. a) first moment and b) square root of the second moment of the boundary disorientation angle distribution, c) mean boundary mobility, d) mean boundary energy. Time is normalized by $1/\left(\mu_c\gamma_c\rho_{0}\right)$. Points are calculated from the list of boundary properties in the Vertex simulation, while lines are the mean-field predictions. The lines are calculated at the means of the reference disorientation angle intervals indicated on plot d).} 
\label{fig:indiv-cells}
\end{figure}

The same analysis is conducted in \autoref{fig:indiv-cells} for the boundary properties of individual cells as a function of their reference disorientation angle $\omega$. 
Again, the evolutions simulated by the full-field model are generally well reproduced by the mean-field model. \autoref{fig:indiv-cells}a shows that the mean boundary disorientation evolution flattens for cells with a large reference disorientation angle $\omega$. This is explained by the fact that the second moment of the cell boundary distribution becomes less sensitive to the reference disorientation
distribution as its own reference disorientation increases (see \autoref{eq:rawsmom-indiv2}). One can also notice, as in the previous figure, the strong relation between the evolution of the first and second moments of boundary disorientation (in \autoref{fig:indiv-cells}a and b). 




Overall, the non-monotonic evolution of boundary disorientation angles induces similar trends in the boundary energies and mobilities. \autoref{fig:indiv-cells}c shows that the mean boundary mobility for orientations with small and large reference disorientation angles $\omega$ are well captured by the mean-field model while the mobility of those with intermediate angles is less well predicted. In \autoref{fig:indiv-cells}d, the mean boundary energy are well predicted for the full range of reference disorientation angle, with again larger discrepancies for cells of intermediate disorientation angle.

\section{Discussion}

\subsection{Comments on the prediction of recrystallization kinetics}

\autoref{fig:kinetics} has shown that the prediction of recrystallization kinetics by the mean-field model is particularly sensitive to the definition of boundary properties. Kinetics are overpredicted when considering only the mean boundary disorientation angles to calculate the mean boundary mobilities and energies, in agreement with the previous attempt of Hurley and Humphreys \cite{hurley_application_2003}. The mean-field model prediction reaches a good agreement with the full-field simulation only by including the contribution of the variances of the boundary disorientation angle distributions. This improvement is due to the assumed  boundary mobility and energy laws. Indeed, the second derivatives of the boundary energy and mobility laws used for calculating the 2$^\text{nd}$ order terms are mostly negative as a function of the disorientation angle ($\gamma''\left(\theta\right)$ is negative for $0^\circ<\theta<15^\circ$ and null above, $\mu''\left(\theta\right)$ is negative above $\sim9^\circ$, see \autoref{app:2derivative}), thus reducing the predicted mean boundary properties and growth rates for the majority of the recrystallized grains. 

Most mean-field models of recrystallization are known to strongly overpredict the density of recrystallized grains. Making the assumption of a site saturation of recrystallized grains, Hurley and Humphreys have reported ratios of 2 to 3 between their model's prediction and experimental measurements at 50\% recrystallization \cite{hurley_application_2003}. In models relying on the Bailey-Hirsch criterion, the overprediction of recrystallized grain density is often hidden by assuming that only
a fraction of the potential recrystallized grains
actually nucleates. This is obtained either by multiplying the Bailey Hirsch criterion itself \cite{beltran_mean_2015} or the number of subgrains meeting the Bailey-Hirsch criterion by fitting constants \cite{zurob_quantitative_2006,dunlop_modelling_2007,lefevre-schlick_activation_2009}. The present results suggest that accounting for the distribution of boundary properties and its time evolution provides a more physical solution to this problem.


One may remark, however, that the complete mean-field model still exhibits discrepancies in predicting the recrystallized grain density (solid black line in \autoref{fig:rhorx}a). This may be caused by assumptions regarding the calculation of growth rates and boundary properties, but it may also be inherent to the mean-field formulation. As already suggested by Zurob \emph{et al.} \cite{zurob_quantitative_2006}, the first recrystallized grains are likely to arise from locations where the energy, and thus the driving force, is higher than the average. Following this argument, mean-field models should have a tendency to underpredict the time at which the first recrystallized grains appear. With the progress of recrystallization and the growth of grains, this effect should become less significant.


\subsection{Boundary dynamics during annealing}

\begin{figure}[htbp]
\centering
\subfigure[]
{\includegraphics[width=0.4\textwidth]{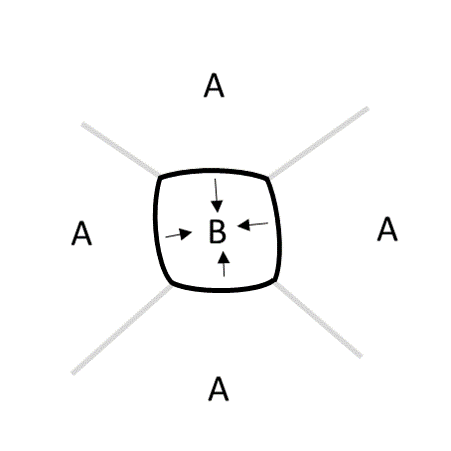}}
\subfigure[]
{\includegraphics[width=0.4\textwidth]{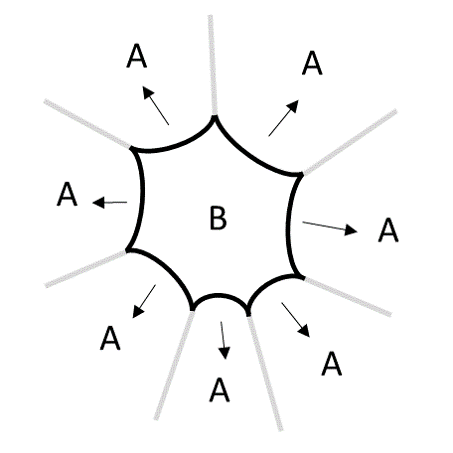}}
\caption{Schematic microstructure of A and B subgrains, with a) a small B subgrain shrinking, b) a large B subgrain growing. High angle boundaries separate the A and B subgrains, while low angles separate subgrains of the same population.}
\label{fig:AB}
\end{figure}

One can understand the time-evolution of boundary properties shown in \autoref{fig:global-spread} and \autoref{fig:indiv-cells} by considering a schematic microstructure composed of A and B subgrains. The A subgrains form the largest fraction of the microstructure, while the B subgrains possess orientations which are far from the average. \autoref{fig:AB}a illustrate the case of a small B subgrain embedded in an environment of A subgrains. Due to its size and its high angle boundaries, the B subgrain shrinks and disapears, inducing a decrease in boundary disorientation angles associated with the A subgrains. By extension, it also slows down the average subgrain growth rates of the A subgrains and prevents them from growing beyond their first neighbour and reaching the critical recrystallized grain size. This evolution is analogous to the concept of orientation pinning sometimes invoked to explain texture developement during recrystallization of aluminium alloys \cite{engler_influence_1998,jensen_orientation_1998}.

By contrast, in \autoref{fig:AB}b the B subgrain is large enough (and has enough neighbours) to grow at the expense of the A subgrains. As the environment of A subgrains remains, the boundary properties of the B subgrain do not change with time. Due to the inverse relation between growth rate and subgrain radius (\autoref{eq:growth-rate2}), the shrinkage of small subgrains with high angle boundaries dominates the microstructure evolution during the early time of annealing (\autoref{fig:AB}a). Once these grains have disapeared, the large subgrains grow and may turn into recrystallized grains surrounded by high angle boundaries (\autoref{fig:AB}b). The improved prediction of the recrystallized grain orientations in the complete mean-field model (\autoref{fig:PF}c) results from taking account of this non-monotonic boundary dynamics. One can finally remark that reviews often make an explicit relation between the onset of recrystallization and the development of high angle boundaries \cite{doherty_current_1997,brechet_nucleation_2006,huang_review_2016}; in the present model, this situation results naturally from the dynamics of subgrains in contact with high angle boundaries.

\subsection{Possible effects of orientation spatial correlations}

\begin{figure}[htbp]
\centering
\subfigure[]
{\includegraphics[width=0.49\textwidth]{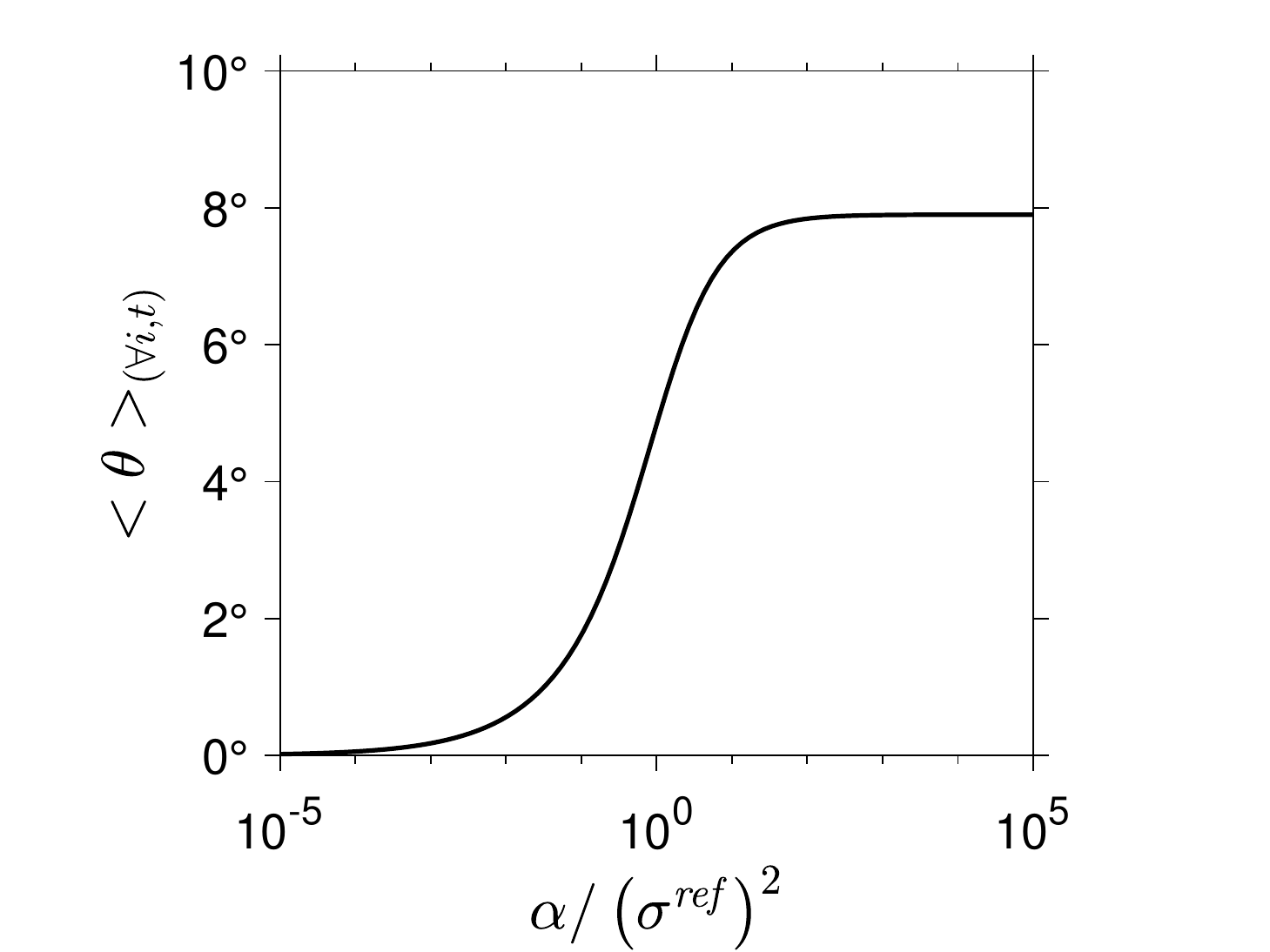}}
\subfigure[]
{\includegraphics[width=0.49\textwidth]{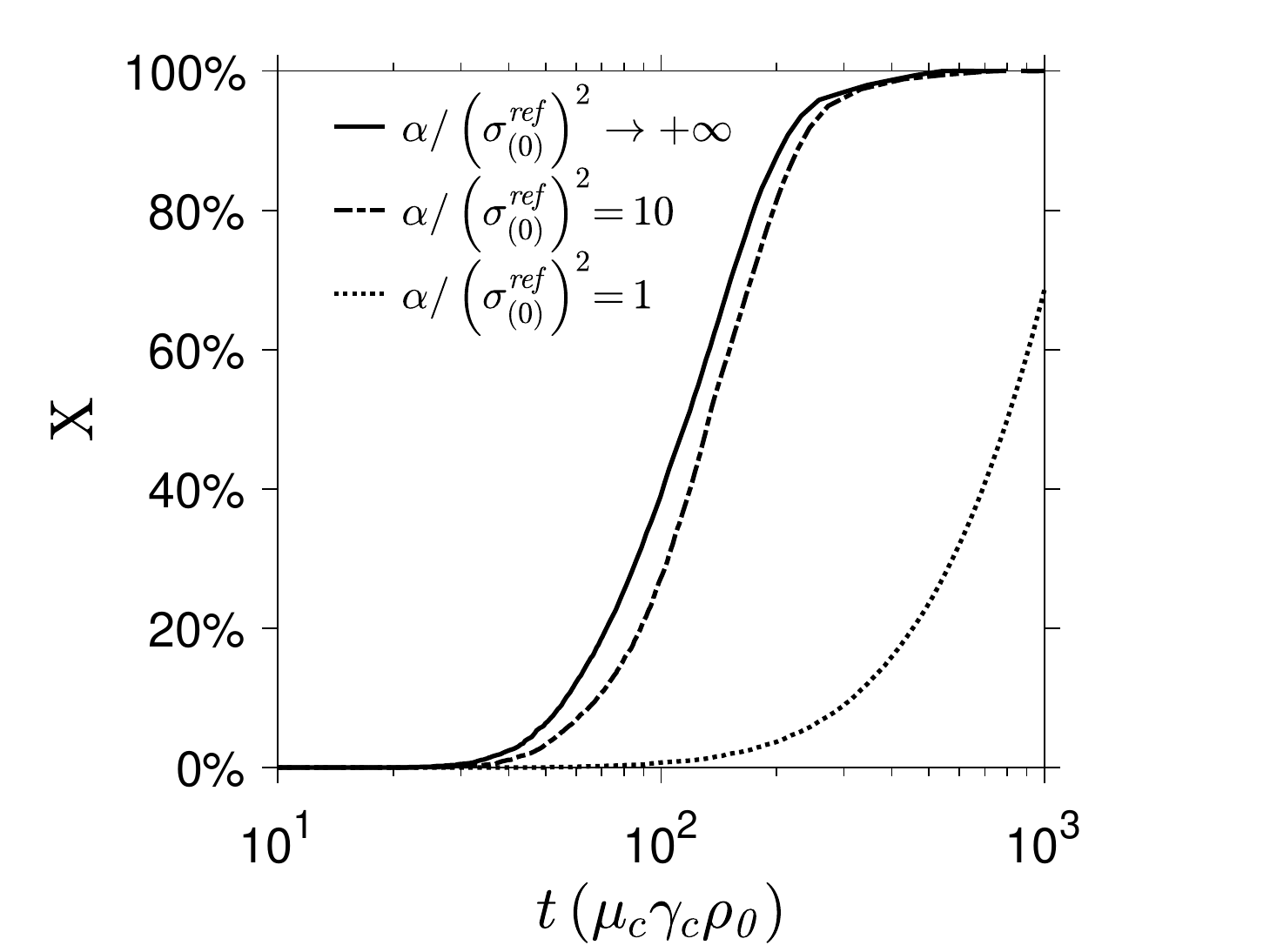}}\caption{a) First moment of the boundary disorientation angle distribution $<\theta>_{\left(\forall i,t\right)}$ as a function of $\alpha$.  b) Effect of $\alpha$ on the prediction of recrystallization kinetics for the microstructure shown earlier. }
\label{fig:Alpha}
\end{figure}

In the mean-field model presented above, spatial correlations between orientations have been introduced through the parameter $\alpha$, which expresses the probability for grains and subgrains to share similar orientations with their neighbours \cite{zecevic_modelling_2019}. As it is fixed for the whole microstructure, it cannot take account of large scale heterogeneities, like those found at deformed grain boundaries or shear bands. \autoref{fig:Alpha}a shows that as $\alpha$ increases, i.e. as correlations vanish, the mean boundary disorientation angle increases towards a constant value. The synthetic full-field microstructures presented above were constructed to have no spatial correlations in the initial state, thus motivating $\alpha$ to be $+\infty$. \autoref{fig:Alpha}b shows that selecting lower values of $\alpha$ would slow down the recrystallization kinetics, as can be expected from the simultaneous decrease in mean boundary disorientation angle.


\begin{figure}[htbp]
\centering
\includegraphics[width=0.49\textwidth]{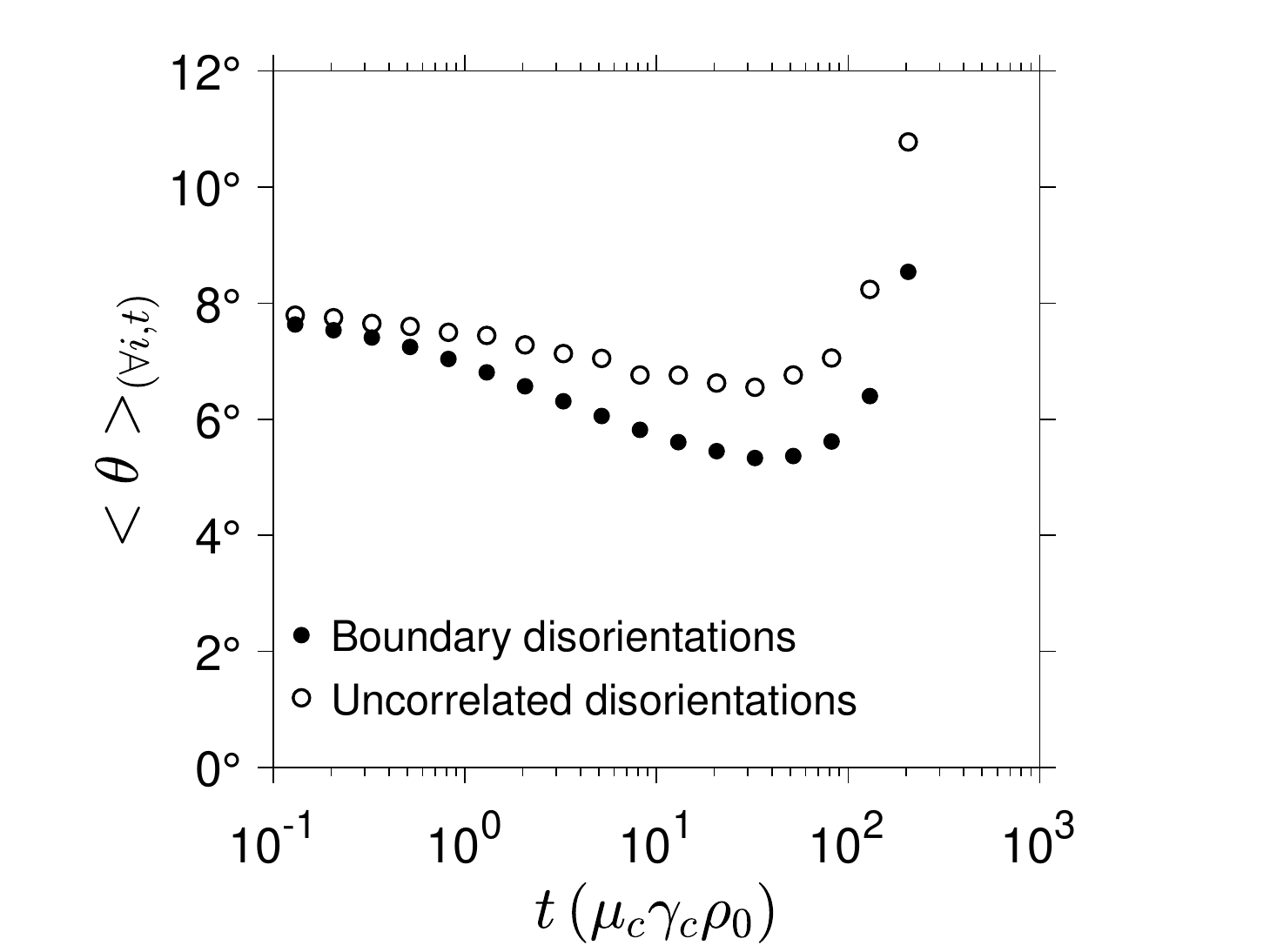}
\caption{Comparison of the mean boundary disorientation angle with the mean uncorrelated disorientation angle calculated from the Vertex microstructure.}
\label{fig:correl}
\end{figure}

It is interesting to notice that spatial correlations may develop even if there are none at the initial state. Orientation pinning is one example where the local configuration of the microstructure affects the evolution of boundary properties. In particular, the schematic in \autoref{fig:AB}a implies that the second nearest neigbour of the A growing subgrains must also be A subgrains, otherwhise the B central subgrain cannot be surrounded by the high angle boundaries leading to its shrinkage. As shown by \autoref{fig:correl}, this scenario is supported by measurements on the Vertex microstructure. In this figure, the mean boundary disorientation angle, measured from the list of boundary disorientation angles in the Vertex microstructure, is compared to the mean uncorrelated disorientation angle. To account for differences in cell size, the mean uncorrelated disorientation angle is calculated as follows: i) each cell is paired with another cell selected at random, ii) the number of pairs per cell is set proportional to its perimeter, and iii) the mean uncorrelated disorientation angle is calculated as the mean disorientation angle between all the pairs. The deviation at long times between the mean boundary disorientation angle and the mean uncorrelated disorientation angle indicates the development of spatial correlations, in a way that could be accounted for by the parameter $\alpha$. The orientation spread remains in any case the most important factor for determining the boundary properties, but the influence of building spatial correlations could become stronger with more heterogeneous microstructures.


%

\subsection{Applicability of the mean-field model}

As a concluding remark, we emphasize that the mean-field model has been constructed with the aim to make it applicable to experimental cases. Orientations spreads \cite{glez_orientation_2001,pantleon_retrieving_2005,krog-pedersen_quantitative_2009,bachmann_inferential_2010}, subgrain sizes \cite{glez_orientation_2001,humphreys_review_2001} and initial spatial correlations \cite{pantleon_correlations_2000,borbely_spatial_2007} can all be measured using EBSD for example. These parameters can also be estimated from crystal-plasticity simulations \cite{kestens_modeling_1996,wenk_deformation-based_1997,zecevic_modelling_2019}. Future implementations could introduce effects from other heterogeneities (e.g. shear bands, transition bands or deformed grain boundaries) following approches adopted elsewhere \cite{pantleon_retrieving_2005,lefevre-schlick_activation_2009}. Further, adopting this to the prediction of recrystallized grains in polycrystals requires an additional scheme to account for the competititve growth between recrystallized grains coming from different parent deformed grains. A first order approach would be to consider the microstructure as a composite of several sub-regions evolving independently, as implemented in previous models for predicting recrystallization kinetics of heterogeneous materials \cite{rollett_computer_1989,kuhbach_statistical_2016}.


\section{Conclusion}

A mean-field model was developed to simulate the time evolution of microstructures during static recrystallization. This model essentially simulates the growth of a population of subgrains contained in well recovered deformed grains, and identifies subgrains above a size threshold as recrystallized grains. At each time increment, the subgrain growth rates are calculated from classical cellular growth laws. The mean subgrain boundary energy and mobility are estimated statistically from knowledge of the orientation spread and of potential spatial correlations between orientation. The orientation spreads considered in this paper are not far from experimental measurements. The model input can be obtained from experimental or synthetic microstructures.

The mean-field model presented here allows one to predict at the same time the recrystallization kinetics and recrystallized grain orientations. The results highlight the significant contribution of the orientation spread and its time-evolution to the determination of boundary properties, the progress of recrystallization and the selection of recrystallized grain orientations. Future work is underway to compare the mean-field model predictions to experimental data.

\section*{Acknowledgements}

The authors express their thanks to Jean-Denis Mithieux and Francis Chassagne for numerous and fruitful discussions on the topic of recrystallization. $\copyright$ Her Majesty the Queen in Right of Canada, as represented by the Minister of Natural Resources, 2020.

%
%
%

\bibliographystyle{elsarticle-num}
\bibliography{Biblio}

\begin{thebibliography}{10}
\expandafter\ifx\csname url\endcsname\relax
  \def\url#1{\texttt{#1}}\fi
\expandafter\ifx\csname urlprefix\endcsname\relax\def\urlprefix{URL }\fi
\expandafter\ifx\csname href\endcsname\relax
  \def\href#1#2{#2} \def\path#1{#1}\fi

\bibitem{bailey_electron_1960}
J.~E. Bailey,
  \href{http://www.tandfonline.com/doi/abs/10.1080/14786436008241221}{Electron
  microscope observations on the annealing processes occurring in cold-worked
  silver}, Philosophical Magazine 5~(56) (1960) 833--842.
\newblock \href {https://doi.org/10.1080/14786436008241221}
  {\path{doi:10.1080/14786436008241221}}.
\newline\urlprefix\url{http://www.tandfonline.com/doi/abs/10.1080/14786436008241221}

\bibitem{bailey_recrystallization_1962}
J.~E. Bailey, P.~B. Hirsch,
  \href{http://rspa.royalsocietypublishing.org/content/267/1328/11}{The
  {Recrystallization} {Process} in {Some} {Polycrystalline} {Metals}},
  Proceedings of the Royal Society of London A: Mathematical, Physical and
  Engineering Sciences 267~(1328) (1962) 11--30.
\newblock \href {https://doi.org/10.1098/rspa.1962.0080}
  {\path{doi:10.1098/rspa.1962.0080}}.
\newline\urlprefix\url{http://rspa.royalsocietypublishing.org/content/267/1328/11}

\bibitem{weygand_nucleation_2000}
D.~Weygand, Y.~Brechett, J.~Lépinoux,
  \href{http://dx.doi.org/10.1080/13642810008216521}{On the nucleation of
  recrystallization by a bulging mechanism: {A} two-dimensional vertex
  simulation}, Philosophical Magazine Part B 80~(11) (2000) 1987--1996.
\newblock \href {https://doi.org/10.1080/13642810008216521}
  {\path{doi:10.1080/13642810008216521}}.
\newline\urlprefix\url{http://dx.doi.org/10.1080/13642810008216521}

\bibitem{holm_abnormal_2003}
E.~A. Holm, M.~A. Miodownik, A.~D. Rollett,
  \href{http://www.sciencedirect.com/science/article/pii/S135964540300079X}{On
  abnormal subgrain growth and the origin of recrystallization nuclei}, Acta
  Materialia 51~(9) (2003) 2701--2716.
\newblock \href {https://doi.org/10.1016/S1359-6454(03)00079-X}
  {\path{doi:10.1016/S1359-6454(03)00079-X}}.
\newline\urlprefix\url{http://www.sciencedirect.com/science/article/pii/S135964540300079X}

\bibitem{hurley_modelling_2003}
P.~J. Hurley, F.~J. Humphreys,
  \href{http://www.sciencedirect.com/science/article/pii/S1359645403001927}{Modelling
  the recrystallization of single-phase aluminium}, Acta Materialia 51~(13)
  (2003) 3779--3793.
\newblock \href {https://doi.org/10.1016/S1359-6454(03)00192-7}
  {\path{doi:10.1016/S1359-6454(03)00192-7}}.
\newline\urlprefix\url{http://www.sciencedirect.com/science/article/pii/S1359645403001927}

\bibitem{zurob_quantitative_2006}
H.~S. Zurob, Y.~Bréchet, J.~Dunlop,
  \href{http://www.sciencedirect.com/science/article/pii/S1359645406003181}{Quantitative
  criterion for recrystallization nucleation in single-phase alloys:
  {Prediction} of critical strains and incubation times}, Acta Materialia
  54~(15) (2006) 3983--3990.
\newblock \href {https://doi.org/10.1016/j.actamat.2006.04.028}
  {\path{doi:10.1016/j.actamat.2006.04.028}}.
\newline\urlprefix\url{http://www.sciencedirect.com/science/article/pii/S1359645406003181}

\bibitem{wang_modeling_2011}
S.~Wang, E.~A. Holm, J.~Suni, M.~H. Alvi, P.~N. Kalu, A.~D. Rollett,
  \href{http://www.sciencedirect.com/science/article/pii/S1359645411001583}{Modeling
  the recrystallized grain size in single phase materials}, Acta Materialia
  59~(10) (2011) 3872--3882.
\newblock \href {https://doi.org/10.1016/j.actamat.2011.03.011}
  {\path{doi:10.1016/j.actamat.2011.03.011}}.
\newline\urlprefix\url{http://www.sciencedirect.com/science/article/pii/S1359645411001583}

\bibitem{favre_nucleation_2013}
J.~Favre, D.~Fabrègue, A.~Chiba, Y.~Bréchet,
  \href{https://doi.org/10.1080/09500839.2013.833352}{Nucleation of
  recrystallization in fine-grained materials: an extension of the
  {Bailey}–{Hirsch} criterion}, Philosophical Magazine Letters 93~(11) (2013)
  631--639.
\newblock \href {https://doi.org/10.1080/09500839.2013.833352}
  {\path{doi:10.1080/09500839.2013.833352}}.
\newline\urlprefix\url{https://doi.org/10.1080/09500839.2013.833352}

\bibitem{huang_review_2016}
K.~Huang, R.~E. Logé,
  \href{http://www.sciencedirect.com/science/article/pii/S0264127516311753}{A
  review of dynamic recrystallization phenomena in metallic materials},
  Materials \& Design 111 (2016) 548--574.
\newblock \href {https://doi.org/10.1016/j.matdes.2016.09.012}
  {\path{doi:10.1016/j.matdes.2016.09.012}}.
\newline\urlprefix\url{http://www.sciencedirect.com/science/article/pii/S0264127516311753}

\bibitem{miesen_highly_2017}
C.~Mießen, N.~Velinov, G.~Gottstein, L.~A. Barrales-Mora,
  \href{https://doi.org/10.1088%2F1361-651x%2Faa8676}{A highly efficient {3D}
  level-set grain growth algorithm tailored for {ccNUMA} architecture},
  Modelling and Simulation in Materials Science and Engineering 25~(8) (2017)
  084002, publisher: IOP Publishing.
\newblock \href {https://doi.org/10.1088/1361-651X/aa8676}
  {\path{doi:10.1088/1361-651X/aa8676}}.
\newline\urlprefix\url{https://doi.org/10.1088%2F1361-651x%2Faa8676}

\bibitem{miyoshi_ultra-large-scale_2017}
E.~Miyoshi, T.~Takaki, M.~Ohno, Y.~Shibuta, S.~Sakane, T.~Shimokawabe, T.~Aoki,
  \href{http://www.nature.com/articles/s41524-017-0029-8}{Ultra-large-scale
  phase-field simulation study of ideal grain growth}, npj Computational
  Materials 3~(1) (2017) 1--6, number: 1 Publisher: Nature Publishing Group.
\newblock \href {https://doi.org/10.1038/s41524-017-0029-8}
  {\path{doi:10.1038/s41524-017-0029-8}}.
\newline\urlprefix\url{http://www.nature.com/articles/s41524-017-0029-8}

\bibitem{suwa_phase-field_2008}
Y.~Suwa, Y.~Saito, H.~Onodera,
  \href{http://www.sciencedirect.com/science/article/pii/S0927025608001699}{Phase-field
  simulation of recrystallization based on the unified subgrain growth theory},
  Computational Materials Science 44~(2) (2008) 286--295.
\newblock \href {https://doi.org/10.1016/j.commatsci.2008.03.025}
  {\path{doi:10.1016/j.commatsci.2008.03.025}}.
\newline\urlprefix\url{http://www.sciencedirect.com/science/article/pii/S0927025608001699}

\bibitem{humphreys_unified_1997}
F.~J. Humphreys,
  \href{http://www.sciencedirect.com/science/article/pii/S1359645497000700}{A
  unified theory of recovery, recrystallization and grain growth, based on the
  stability and growth of cellular microstructures—{I}. {The} basic model},
  Acta Materialia 45~(10) (1997) 4231--4240.
\newblock \href {https://doi.org/10.1016/S1359-6454(97)00070-0}
  {\path{doi:10.1016/S1359-6454(97)00070-0}}.
\newline\urlprefix\url{http://www.sciencedirect.com/science/article/pii/S1359645497000700}

\bibitem{rollett_growth_1997}
A.~Rollett,
  \href{https://linkinghub.elsevier.com/retrieve/pii/S1359646296005015}{On the
  growth of abnormal grains}, Scripta Materialia 36~(9) (1997) 975--980.
\newblock \href {https://doi.org/10.1016/S1359-6462(96)00501-5}
  {\path{doi:10.1016/S1359-6462(96)00501-5}}.
\newline\urlprefix\url{https://linkinghub.elsevier.com/retrieve/pii/S1359646296005015}

\bibitem{razzak_simple_2012}
M.~A. Razzak, M.~Perez, T.~Sourmail, S.~Cazottes, M.~Frotey, A {Simple} {Model}
  for {Abnormal} {Grain} {Growth}, ISIJ International 52~(12) (2012)
  2278--2282.
\newblock \href {https://doi.org/10.2355/isijinternational.52.2278}
  {\path{doi:10.2355/isijinternational.52.2278}}.

\bibitem{syha_conditions_2012}
M.~Syha, D.~Weygand,
  \href{https://www.scientific.net/MSF.715-716.563}{Conditions for the
  {Occurrence} of {Abnormal} {Grain} {Growth} {Studied} by a 3 {D} {Vertex}
  {Dynamics} {Model}}, 2012.
\newblock \href {https://doi.org/10.4028/www.scientific.net/MSF.715-716.563}
  {\path{doi:10.4028/www.scientific.net/MSF.715-716.563}}.
\newline\urlprefix\url{https://www.scientific.net/MSF.715-716.563}

\bibitem{dunlop_modelling_2007}
J.~W.~C. Dunlop, Y.~J.~M. Bréchet, L.~Legras, H.~S. Zurob,
  \href{http://www.sciencedirect.com/science/article/pii/S0022311507000104}{Modelling
  isothermal and non-isothermal recrystallisation kinetics: {Application} to
  {Zircaloy}-4}, Journal of Nuclear Materials 366~(1) (2007) 178--186.
\newblock \href {https://doi.org/10.1016/j.jnucmat.2006.12.074}
  {\path{doi:10.1016/j.jnucmat.2006.12.074}}.
\newline\urlprefix\url{http://www.sciencedirect.com/science/article/pii/S0022311507000104}

\bibitem{beltran_mean_2015}
O.~Beltran, K.~Huang, R.~E. Logé,
  \href{http://www.sciencedirect.com/science/article/pii/S0927025615001445}{A
  mean field model of dynamic and post-dynamic recrystallization predicting
  kinetics, grain size and flow stress}, Computational Materials Science 102
  (2015) 293--303.
\newblock \href {https://doi.org/10.1016/j.commatsci.2015.02.043}
  {\path{doi:10.1016/j.commatsci.2015.02.043}}.
\newline\urlprefix\url{http://www.sciencedirect.com/science/article/pii/S0927025615001445}

\bibitem{kestens_modeling_1996}
L.~Kestens, J.~J. Jonas, \href{https://doi.org/10.1007/BF02647756}{Modeling
  texture change during the static recrystallization of interstitial free
  steels}, Metallurgical and Materials Transactions A 27~(1) (1996) 155--164.
\newblock \href {https://doi.org/10.1007/BF02647756}
  {\path{doi:10.1007/BF02647756}}.
\newline\urlprefix\url{https://doi.org/10.1007/BF02647756}

\bibitem{wenk_deformation-based_1997}
H.~R. Wenk, G.~Canova, Y.~Bréchet, L.~Flandin,
  \href{http://www.sciencedirect.com/science/article/pii/S1359645496004090}{A
  deformation-based model for recrystallization of anisotropic materials}, Acta
  Materialia 45~(8) (1997) 3283--3296.
\newblock \href {https://doi.org/10.1016/S1359-6454(96)00409-0}
  {\path{doi:10.1016/S1359-6454(96)00409-0}}.
\newline\urlprefix\url{http://www.sciencedirect.com/science/article/pii/S1359645496004090}

\bibitem{zecevic_modelling_2019}
M.~Zecevic, R.~A. Lebensohn, R.~J. McCabe, M.~Knezevic,
  \href{http://www.sciencedirect.com/science/article/pii/S1359645418308772}{Modelling
  recrystallization textures driven by intragranular fluctuations implemented
  in the viscoplastic self-consistent formulation}, Acta Materialia 164 (2019)
  530--546.
\newblock \href {https://doi.org/10.1016/j.actamat.2018.11.002}
  {\path{doi:10.1016/j.actamat.2018.11.002}}.
\newline\urlprefix\url{http://www.sciencedirect.com/science/article/pii/S1359645418308772}

\bibitem{weygand_vertex_1998}
D.~Weygand, Y.~Bréchet, J.~Lépinoux,
  \href{http://dx.doi.org/10.1080/13642819808206731}{A vertex dynamics
  simulation of grain growth in two dimensions}, Philosophical Magazine Part B
  78~(4) (1998) 329--352.
\newblock \href {https://doi.org/10.1080/13642819808206731}
  {\path{doi:10.1080/13642819808206731}}.
\newline\urlprefix\url{http://dx.doi.org/10.1080/13642819808206731}

\bibitem{piekos_generalized_2008}
K.~Piekos, J.~Tarasiuk, K.~Wierzbanowski, B.~Bacroix,
  \href{http://www.sciencedirect.com/science/article/pii/S0927025607002819}{Generalized
  vertex model of recrystallization – {Application} to polycrystalline
  copper}, Computational Materials Science 42~(4) (2008) 584--594.
\newblock \href {https://doi.org/10.1016/j.commatsci.2007.09.014}
  {\path{doi:10.1016/j.commatsci.2007.09.014}}.
\newline\urlprefix\url{http://www.sciencedirect.com/science/article/pii/S0927025607002819}

\bibitem{mellbin_combined_2015}
Y.~Mellbin, H.~Hallberg, M.~Ristinmaa,
  \href{http://stacks.iop.org/0965-0393/23/i=4/a=045011?key=crossref.4667c57f5766b56935eeb38b10ca0c5c}{A
  combined crystal plasticity and graph-based vertex model of dynamic
  recrystallization at large deformations}, Modelling and Simulation in
  Materials Science and Engineering 23~(4) (2015) 045011.
\newblock \href {https://doi.org/10.1088/0965-0393/23/4/045011}
  {\path{doi:10.1088/0965-0393/23/4/045011}}.
\newline\urlprefix\url{http://stacks.iop.org/0965-0393/23/i=4/a=045011?key=crossref.4667c57f5766b56935eeb38b10ca0c5c}

\bibitem{huang_subgrain_2000}
Y.~Huang, F.~J. Humphreys,
  \href{http://www.sciencedirect.com/science/article/pii/S1359645499004188}{Subgrain
  growth and low angle boundary mobility in aluminium crystals of orientation
  \{110\}(001)}, Acta Materialia 48~(8) (2000) 2017--2030.
\newblock \href {https://doi.org/10.1016/S1359-6454(99)00418-8}
  {\path{doi:10.1016/S1359-6454(99)00418-8}}.
\newline\urlprefix\url{http://www.sciencedirect.com/science/article/pii/S1359645499004188}

\bibitem{engler_introduction_2009}
O.~Engler, V.~Randle, Introduction to {Texture} {Analysis}: {Macrotexture},
  {Microtexture}, and {Orientation} {Mapping}, 2nd {Edition}, CRC Press, 2009.

\bibitem{read_dislocation_1950}
W.~T. Read, W.~Shockley,
  \href{http://link.aps.org/doi/10.1103/PhysRev.78.275}{Dislocation {Models} of
  {Crystal} {Grain} {Boundaries}}, Phys. Rev. 78~(3) (1950) 275--289.
\newblock \href {https://doi.org/10.1103/PhysRev.78.275}
  {\path{doi:10.1103/PhysRev.78.275}}.
\newline\urlprefix\url{http://link.aps.org/doi/10.1103/PhysRev.78.275}

\bibitem{louat_theory_1974}
N.~P. Louat,
  \href{http://www.sciencedirect.com/science/article/pii/0001616074900819}{On
  the theory of normal grain growth}, Acta Metallurgica 22~(6) (1974) 721--724.
\newblock \href {https://doi.org/10.1016/0001-6160(74)90081-9}
  {\path{doi:10.1016/0001-6160(74)90081-9}}.
\newline\urlprefix\url{http://www.sciencedirect.com/science/article/pii/0001616074900819}

\bibitem{srolovitz_computer_1984}
D.~J. Srolovitz, M.~P. Anderson, P.~S. Sahni, G.~S. Grest,
  \href{http://www.sciencedirect.com/science/article/pii/0001616084901524}{Computer
  simulation of grain growth—{II}. {Grain} size distribution, topology, and
  local dynamics}, Acta Metallurgica 32~(5) (1984) 793--802.
\newblock \href {https://doi.org/10.1016/0001-6160(84)90152-4}
  {\path{doi:10.1016/0001-6160(84)90152-4}}.
\newline\urlprefix\url{http://www.sciencedirect.com/science/article/pii/0001616084901524}

\bibitem{glez_orientation_2001}
J.~C. Glez, J.~Driver,
  \href{http://scripts.iucr.org/cgi-bin/paper?vi0144}{Orientation distribution
  analysis in deformed grains}, J Appl Cryst, J Appl Crystallogr 34~(3) (2001)
  280--288.
\newblock \href {https://doi.org/10.1107/S0021889801003077}
  {\path{doi:10.1107/S0021889801003077}}.
\newline\urlprefix\url{http://scripts.iucr.org/cgi-bin/paper?vi0144}

\bibitem{pantleon_retrieving_2005}
W.~Pantleon,
  \href{http://www.tandfonline.com/doi/full/10.1179/174328405X71657}{Retrieving
  orientation correlations in deformation structures from orientation maps},
  Materials Science and Technology 21~(12) (2005) 1392--1396.
\newblock \href {https://doi.org/10.1179/174328405X71657}
  {\path{doi:10.1179/174328405X71657}}.
\newline\urlprefix\url{http://www.tandfonline.com/doi/full/10.1179/174328405X71657}

\bibitem{miodownik_mark_a._scaling_2001}
{Miodownik Mark A.}, {Smereka Peter}, {Srolovitz David J.}, {Holm Elizabeth
  A.},
  \href{https://royalsocietypublishing.org/doi/abs/10.1098/rspa.2001.0794}{Scaling
  of dislocation cell structures: diffusion in orientation space}, Proceedings
  of the Royal Society of London. Series A: Mathematical, Physical and
  Engineering Sciences 457~(2012) (2001) 1807--1819.
\newblock \href {https://doi.org/10.1098/rspa.2001.0794}
  {\path{doi:10.1098/rspa.2001.0794}}.
\newline\urlprefix\url{https://royalsocietypublishing.org/doi/abs/10.1098/rspa.2001.0794}

\bibitem{pantleon_dislocation_2001}
W.~Pantleon, N.~Hansen,
  \href{http://www.sciencedirect.com/science/article/pii/S1359645401000271}{Dislocation
  boundaries—the distribution function of disorientation angles}, Acta
  Materialia 49~(8) (2001) 1479--1493.
\newblock \href {https://doi.org/10.1016/S1359-6454(01)00027-1}
  {\path{doi:10.1016/S1359-6454(01)00027-1}}.
\newline\urlprefix\url{http://www.sciencedirect.com/science/article/pii/S1359645401000271}

\bibitem{hughes_scaling_1998}
D.~A. Hughes, D.~C. Chrzan, Q.~Liu, N.~Hansen,
  \href{https://link.aps.org/doi/10.1103/PhysRevLett.81.4664}{Scaling of
  {Misorientation} {Angle} {Distributions}}, Physical Review Letters 81~(21)
  (1998) 4664--4667.
\newblock \href {https://doi.org/10.1103/PhysRevLett.81.4664}
  {\path{doi:10.1103/PhysRevLett.81.4664}}.
\newline\urlprefix\url{https://link.aps.org/doi/10.1103/PhysRevLett.81.4664}

\bibitem{hillert_theory_1965}
M.~Hillert,
  \href{http://www.sciencedirect.com/science/article/pii/0001616065902002}{On
  the theory of normal and abnormal grain growth}, Acta Metallurgica 13~(3)
  (1965) 227--238.
\newblock \href {https://doi.org/10.1016/0001-6160(65)90200-2}
  {\path{doi:10.1016/0001-6160(65)90200-2}}.
\newline\urlprefix\url{http://www.sciencedirect.com/science/article/pii/0001616065902002}

\bibitem{abbruzzese_theory_1986}
G.~Abbruzzese, K.~Lücke,
  \href{http://www.sciencedirect.com/science/article/pii/0001616086900647}{A
  theory of texture controlled grain growth—{I}. {Derivation} and general
  discussion of the model}, Acta Metallurgica 34~(5) (1986) 905--914.
\newblock \href {https://doi.org/10.1016/0001-6160(86)90064-7}
  {\path{doi:10.1016/0001-6160(86)90064-7}}.
\newline\urlprefix\url{http://www.sciencedirect.com/science/article/pii/0001616086900647}

\bibitem{park_moments_1961}
J.~H. Park,
  \href{http://www.ams.org/qam/1961-19-01/S0033-569X-1961-0119222-9/}{Moments
  of the generalized {Rayleigh} distribution}, Quarterly of Applied Mathematics
  19~(1) (1961) 45--49.
\newblock \href {https://doi.org/10.1090/qam/119222}
  {\path{doi:10.1090/qam/119222}}.
\newline\urlprefix\url{http://www.ams.org/qam/1961-19-01/S0033-569X-1961-0119222-9/}

\bibitem{krog-pedersen_quantitative_2009}
S.~Krog-Pedersen, J.~R. Bowen, W.~Pantleon,
  \href{http://www.hanser-elibrary.com/doi/abs/10.3139/146.110032}{Quantitative
  characterization of the orientation spread within individual grains in copper
  after tensile deformation}, International Journal of Materials Research
  100~(3) (2009) 433--438.
\newblock \href {https://doi.org/10.3139/146.110032}
  {\path{doi:10.3139/146.110032}}.
\newline\urlprefix\url{http://www.hanser-elibrary.com/doi/abs/10.3139/146.110032}

\bibitem{despres_contribution_2020}
A.~Després, M.~Zecevic, R.~A. Lebensohn, J.~D. Mithieux, F.~Chassagne, C.~W.
  Sinclair,
  \href{http://www.sciencedirect.com/science/article/pii/S1359645419306858}{Contribution
  of intragranular misorientations to the cold rolling textures of ferritic
  stainless steels}, Acta Materialia 182 (2020) 184--196.
\newblock \href {https://doi.org/10.1016/j.actamat.2019.10.023}
  {\path{doi:10.1016/j.actamat.2019.10.023}}.
\newline\urlprefix\url{http://www.sciencedirect.com/science/article/pii/S1359645419306858}

\bibitem{hurley_application_2003}
P.~J. Hurley, F.~J. Humphreys,
  \href{http://www.sciencedirect.com/science/article/pii/S135964540200513X}{The
  application of {EBSD} to the study of substructural development in a cold
  rolled single-phase aluminium alloy}, Acta Materialia 51~(4) (2003)
  1087--1102.
\newblock \href {https://doi.org/10.1016/S1359-6454(02)00513-X}
  {\path{doi:10.1016/S1359-6454(02)00513-X}}.
\newline\urlprefix\url{http://www.sciencedirect.com/science/article/pii/S135964540200513X}

\bibitem{perryman_recrystallization_1955}
E.~C.~W. Perryman, Recrystallization {Characteristics} of {Superpurity} {Base}
  {AI}-{Mg} {Alloys} {Containing} 0 to 5 {Pct} {M} g, Trans AIME (1955)
  369--378.

\bibitem{mishin_recovery_2013}
O.~V. Mishin, A.~Godfrey, D.~Juul~Jensen, N.~Hansen,
  \href{http://www.sciencedirect.com/science/article/pii/S1359645413003923}{Recovery
  and recrystallization in commercial purity aluminum cold rolled to an
  ultrahigh strain}, Acta Materialia 61~(14) (2013) 5354--5364.
\newblock \href {https://doi.org/10.1016/j.actamat.2013.05.024}
  {\path{doi:10.1016/j.actamat.2013.05.024}}.
\newline\urlprefix\url{http://www.sciencedirect.com/science/article/pii/S1359645413003923}

\bibitem{lefevre-schlick_activation_2009}
F.~Lefevre-Schlick, Y.~Brechet, H.~S. Zurob, G.~Purdy, D.~Embury,
  \href{http://www.sciencedirect.com/science/article/pii/S0921509308011702}{On
  the activation of recrystallization nucleation sites in {Cu} and {Fe}},
  Materials Science and Engineering: A 502~(1) (2009) 70--78.
\newblock \href {https://doi.org/10.1016/j.msea.2008.10.015}
  {\path{doi:10.1016/j.msea.2008.10.015}}.
\newline\urlprefix\url{http://www.sciencedirect.com/science/article/pii/S0921509308011702}

\bibitem{engler_influence_1998}
O.~Engler,
  \href{http://www.sciencedirect.com/science/article/pii/S1359645497003546}{On
  the influence of orientation pinning on growth selection of
  recrystallisation}, Acta Materialia 46~(5) (1998) 1555--1568.
\newblock \href {https://doi.org/10.1016/S1359-6454(97)00354-6}
  {\path{doi:10.1016/S1359-6454(97)00354-6}}.
\newline\urlprefix\url{http://www.sciencedirect.com/science/article/pii/S1359645497003546}

\bibitem{jensen_orientation_1998}
D.~J. Jensen, K.~Mehnert,
  \href{https://orbit.dtu.dk/en/publications/orientation-pinning-during-growth}{Orientation
  pinning during growth}, in: Grain growth in polycrystalline materials 3,
  Minerals, Metals and Materials Society, 1998, pp. 251--262.
\newline\urlprefix\url{https://orbit.dtu.dk/en/publications/orientation-pinning-during-growth}

\bibitem{doherty_current_1997}
R.~D. Doherty, D.~A. Hughes, F.~J. Humphreys, J.~J. Jonas, D.~Juul~Jensen,
  M.~E. Kassner, W.~E. King, T.~R. McNelley, H.~J. McQueen, A.~D. Rollett,
  \href{http://www.sciencedirect.com/science/article/pii/S0921509397004243}{Current
  issues in recrystallization: a review}, Materials Science and Engineering: A
  238~(2) (1997) 219--274.
\newblock \href {https://doi.org/10.1016/S0921-5093(97)00424-3}
  {\path{doi:10.1016/S0921-5093(97)00424-3}}.
\newline\urlprefix\url{http://www.sciencedirect.com/science/article/pii/S0921509397004243}

\bibitem{brechet_nucleation_2006}
Y.~Bréchet, G.~Martin,
  \href{http://www.sciencedirect.com/science/article/pii/S1631070506002258}{Nucleation
  problems in metallurgy of the solid state: recent developments and open
  questions}, Comptes Rendus de Physique 7~(9–10) (2006) 959--976.
\newblock \href {https://doi.org/10.1016/j.crhy.2006.10.014}
  {\path{doi:10.1016/j.crhy.2006.10.014}}.
\newline\urlprefix\url{http://www.sciencedirect.com/science/article/pii/S1631070506002258}

\bibitem{bachmann_inferential_2010}
F.~Bachmann, R.~Hielscher, P.~E. Jupp, W.~Pantleon, H.~Schaeben, E.~Wegert,
  \href{http://scripts.iucr.org/cgi-bin/paper?cg5145}{Inferential statistics of
  electron backscatter diffraction data from within individual crystalline
  grains}, J Appl Cryst, J Appl Crystallogr 43~(6) (2010) 1338--1355.
\newblock \href {https://doi.org/10.1107/S002188981003027X}
  {\path{doi:10.1107/S002188981003027X}}.
\newline\urlprefix\url{http://scripts.iucr.org/cgi-bin/paper?cg5145}

\bibitem{humphreys_review_2001}
F.~J. Humphreys,
  \href{http://link.springer.com/article/10.1023/A%3A1017973432592}{Review
  {Grain} and subgrain characterisation by electron backscatter diffraction},
  Journal of Materials Science 36~(16) (2001) 3833--3854.
\newblock \href {https://doi.org/10.1023/A:1017973432592}
  {\path{doi:10.1023/A:1017973432592}}.
\newline\urlprefix\url{http://link.springer.com/article/10.1023/A%3A1017973432592}

\bibitem{pantleon_correlations_2000}
W.~Pantleon, D.~Stoyan,
  \href{http://www.sciencedirect.com/science/article/pii/S1359645400000835}{Correlations
  between disorientations in neighbouring dislocation boundaries}, Acta
  Materialia 48~(11) (2000) 3005--3014.
\newblock \href {https://doi.org/10.1016/S1359-6454(00)00083-5}
  {\path{doi:10.1016/S1359-6454(00)00083-5}}.
\newline\urlprefix\url{http://www.sciencedirect.com/science/article/pii/S1359645400000835}

\bibitem{borbely_spatial_2007}
A.~Borbély, C.~Maurice, D.~Piot, J.~H. Driver,
  \href{http://www.sciencedirect.com/science/article/pii/S1359645406006203}{Spatial
  characterisation of the orientation distributions in a stable plane
  strain-compressed {Cu} crystal: {A} statistical analysis}, Acta Materialia
  55~(2) (2007) 487--496.
\newblock \href {https://doi.org/10.1016/j.actamat.2006.08.043}
  {\path{doi:10.1016/j.actamat.2006.08.043}}.
\newline\urlprefix\url{http://www.sciencedirect.com/science/article/pii/S1359645406006203}

\bibitem{rollett_computer_1989}
A.~Rollett, D.~Srolovitz, R.~Doherty, M.~Anderson,
  \href{https://linkinghub.elsevier.com/retrieve/pii/0001616089902472}{Computer
  simulation of recrystallization in non-uniformly deformed metals}, Acta
  Metallurgica 37~(2) (1989) 627--639.
\newblock \href {https://doi.org/10.1016/0001-6160(89)90247-2}
  {\path{doi:10.1016/0001-6160(89)90247-2}}.
\newline\urlprefix\url{https://linkinghub.elsevier.com/retrieve/pii/0001616089902472}

\bibitem{kuhbach_statistical_2016}
M.~Kühbach, G.~Gottstein, L.~A. Barrales-Mora,
  \href{http://www.sciencedirect.com/science/article/pii/S1359645416300659}{A
  statistical ensemble cellular automaton microstructure model for primary
  recrystallization}, Acta Materialia 107 (2016) 366--376.
\newblock \href {https://doi.org/10.1016/j.actamat.2016.01.068}
  {\path{doi:10.1016/j.actamat.2016.01.068}}.
\newline\urlprefix\url{http://www.sciencedirect.com/science/article/pii/S1359645416300659}

\bibitem{macpherson_von_2007}
R.~D. MacPherson, D.~J. Srolovitz,
  \href{https://www-nature-com.ezproxy.library.ubc.ca/articles/nature05745}{The
  von {Neumann} relation generalized to coarsening of three-dimensional
  microstructures}, Nature 446~(7139) (2007) 1053--1055.
\newblock \href {https://doi.org/10.1038/nature05745}
  {\path{doi:10.1038/nature05745}}.
\newline\urlprefix\url{https://www-nature-com.ezproxy.library.ubc.ca/articles/nature05745}

\bibitem{le_generalization_2009}
T.~Le, Q.~Du, \href{https://projecteuclid.org/euclid.cms/1243443993}{A
  generalization of the three-dimensional {MacPherson}-{Srolovitz} formula},
  Communications in Mathematical Sciences 7~(2) (2009) 511--520.
\newline\urlprefix\url{https://projecteuclid.org/euclid.cms/1243443993}

\bibitem{zhang_three-dimensional_2018}
J.~Zhang, Y.~Zhang, W.~Ludwig, D.~Rowenhorst, P.~W. Voorhees, H.~F. Poulsen,
  \href{http://www.sciencedirect.com/science/article/pii/S1359645418304890}{Three-dimensional
  grain growth in pure iron. {Part} {I}. statistics on the grain level}, Acta
  Materialia 156 (2018) 76--85.
\newblock \href {https://doi.org/10.1016/j.actamat.2018.06.021}
  {\path{doi:10.1016/j.actamat.2018.06.021}}.
\newline\urlprefix\url{http://www.sciencedirect.com/science/article/pii/S1359645418304890}

\bibitem{glicksman_mean_2009}
M.~E. Glicksman, P.~R. Rios, D.~J. Lewis,
  \href{https://doi.org/10.1080/14786430802651513}{Mean width and caliper
  characteristics of network polyhedra}, Philosophical Magazine 89~(4) (2009)
  389--403.
\newblock \href {https://doi.org/10.1080/14786430802651513}
  {\path{doi:10.1080/14786430802651513}}.
\newline\urlprefix\url{https://doi.org/10.1080/14786430802651513}

\bibitem{mirevski_fractional_2010}
S.~P. Mirevski, L.~Boyadjiev,
  \href{http://www.sciencedirect.com/science/article/pii/S0898122109004106}{On
  some fractional generalizations of the {Laguerre} polynomials and the
  {Kummer} function}, Computers \& Mathematics with Applications 59~(3) (2010)
  1271--1277.
\newblock \href {https://doi.org/10.1016/j.camwa.2009.06.037}
  {\path{doi:10.1016/j.camwa.2009.06.037}}.
\newline\urlprefix\url{http://www.sciencedirect.com/science/article/pii/S0898122109004106}

\end{thebibliography}

\appendix
\begin{appendices}

\section{Growth equations for 3 dimensional microstructures}
\label{app:3DMullins}

The growth rate of a 3 dimensional cell embedded in a cellular structure of uniform boundary energy and mobility is given by the MacPherson-Srolovitz equation \cite{macpherson_von_2007}:


\begin{equation}
    \frac{dV}{dt}=-2\pi M \Gamma\left(\mathcal{L}-\frac{1}{6}\mathcal{M}\right)
    \label{eq:MacPherson-iso}
\end{equation}

Where $V$ is the cell volume, $M$ and $\Gamma$ are its boundary mobility and energy, $\mathcal{L}$ is so called mean width of the cell, and $\mathcal{M}=\sum_{i=1}^{n_l} l_i$ is the sum of the cell edge length, running over the $n_l$ edges. For simplicity, we omit subscripts associated with time or cell index. This equation assumes that the turning angles at the cell edges are all at equilibrium and equal to $\pi/3$, as in the 2 dimensional case.

To account for heterogeneous boundary properties, the second right term in parenthesis in \autoref{eq:MacPherson-iso} can be replaced by $1/\left(2\pi\right)\sum_{i=1}^{n_l}\xi l_i$, with $\xi$ the equilibrium turning angle measured at the cell edges (see \cite{macpherson_von_2007,le_generalization_2009}). In a mean-field environement, $\xi$ is the same for all cell edges, and the growth rate is given by:

\begin{equation}
    \frac{dV}{dt}=-2\pi M \Gamma\left(\mathcal{L}-\frac{\xi}{2\pi}\mathcal{M}\right)
\label{eq:MacPherson-aniso}
\end{equation}

When all boundaries have equal energy, $\xi=\pi/3$, and \autoref{eq:MacPherson-aniso} reduces to \autoref{eq:MacPherson-iso}.

The mean width $\mathcal{L}$ can be calculated from formulas given in ref. \cite{macpherson_von_2007}. It takes a value of $4R$ for a sphere. It is possible to show that it is strictly above $4R$ for polyhedrons at constant volume, with $R$ the volume equivalent radius. Zhang \emph{et al.} \cite{zhang_three-dimensional_2018} found on a 3D microstructure obtained by diffraction contrast tomography that the grain mean width follows on average:

\begin{equation}
   \mathcal{L}\approx5R 
   \label{eq:L}
\end{equation}

Drawing analogies between the MacPherson-Srolovitz equation and the Hillert equation, they suggested the sum of edges lengths to follow a quadratic relation with the volume equivalent radius. After rerranging the relation proposed in the original publication to make it dependent on the cell radius \cite{zhang_three-dimensional_2018}:

\begin{equation}
   \mathcal{M}\approx6\left(3R+ \frac{16R^2}{9\bar{R}}\right)
   \label{eq:Mterm}
\end{equation}


The work of Glicksman \emph{et al.} \cite{glicksman_mean_2009}, showing for some solids that $\xi\mathcal{M}$ varies roughly linearly with $\xi$, suggests by induction that $\mathcal{M}$ can be considered independent of $\xi$. Finally, inserting \autoref{eq:L} and \autoref{eq:Mterm} in \autoref{eq:MacPherson-aniso}, and expressing growth rate in terms of the cell radius, one obtains:

\begin{equation}
    \frac{dR}{dt}=\frac{M \Gamma}{2R} \left(a\left(3+\frac{16R}{9\bar{R}}\right)-5\right)
\label{eq:MacPherson-aniso2}
\end{equation}

Which is the same form as \autoref{eq:growth-rate2} used for 2 dimensional microstructures. When boundary energy is uniform, $a=1$ and \autoref{eq:MacPherson-aniso2} reduces to the classical Hillert equation for 3 dimensional microstructures \cite{hillert_theory_1965}.

\section{Generalized Laguerre function}
\label{app:Laguerre}

A review of the generalized Laguerre polynomials and functions has been written by Mirevski and Boyadjiev \cite{mirevski_fractional_2010}. Laguerre functions are solutions of the Laguerre differential equation with fractional coefficients. First, the binomial coefficient with real arguments $\alpha$ and $\beta$ is defined as \cite{mirevski_fractional_2010}:

\begin{equation}
    \begin{pmatrix}
\alpha \\
\beta
\end{pmatrix}
=\frac{\Gamma\left(1+\alpha\right)}{\Gamma\left(1+\beta\right)\Gamma\left(1+\alpha-\beta\right)}
\end{equation}

Where $\Gamma$ is, for this particular equation, the gamma function. Laguerre functions are expressed by the series expansion \cite{mirevski_fractional_2010}:



\begin{equation}
L_\nu^{\left(\alpha\right)}\left(x\right)=
\begin{pmatrix}
\nu+\alpha \\
\nu
\end{pmatrix}
\sum_{k=0}^{\infty}\frac{\left(-\nu\right)\left(-\nu+1\right)...\left(-\nu+k-1\right)}{\left(\alpha+1\right)\left(\alpha+2\right)...\left(\alpha+k\right)}
    \frac{\left(-x\right)^k}{k!}
\end{equation}

For $\nu=1/2$ and $\alpha=1/2$, the right term is reduced to:
\begin{equation}
L_{1/2}^{\left(1/2\right)}\left(x\right)=
\begin{pmatrix}
1 \\
1/2
\end{pmatrix}
\sum_{k=0}^{\infty}\frac{1}{1-4k^2}
    \frac{\left(-x\right)^k}{k!}
\end{equation}

\begin{figure}[htbp]
    \centering
    \includegraphics[width=0.5\textwidth]{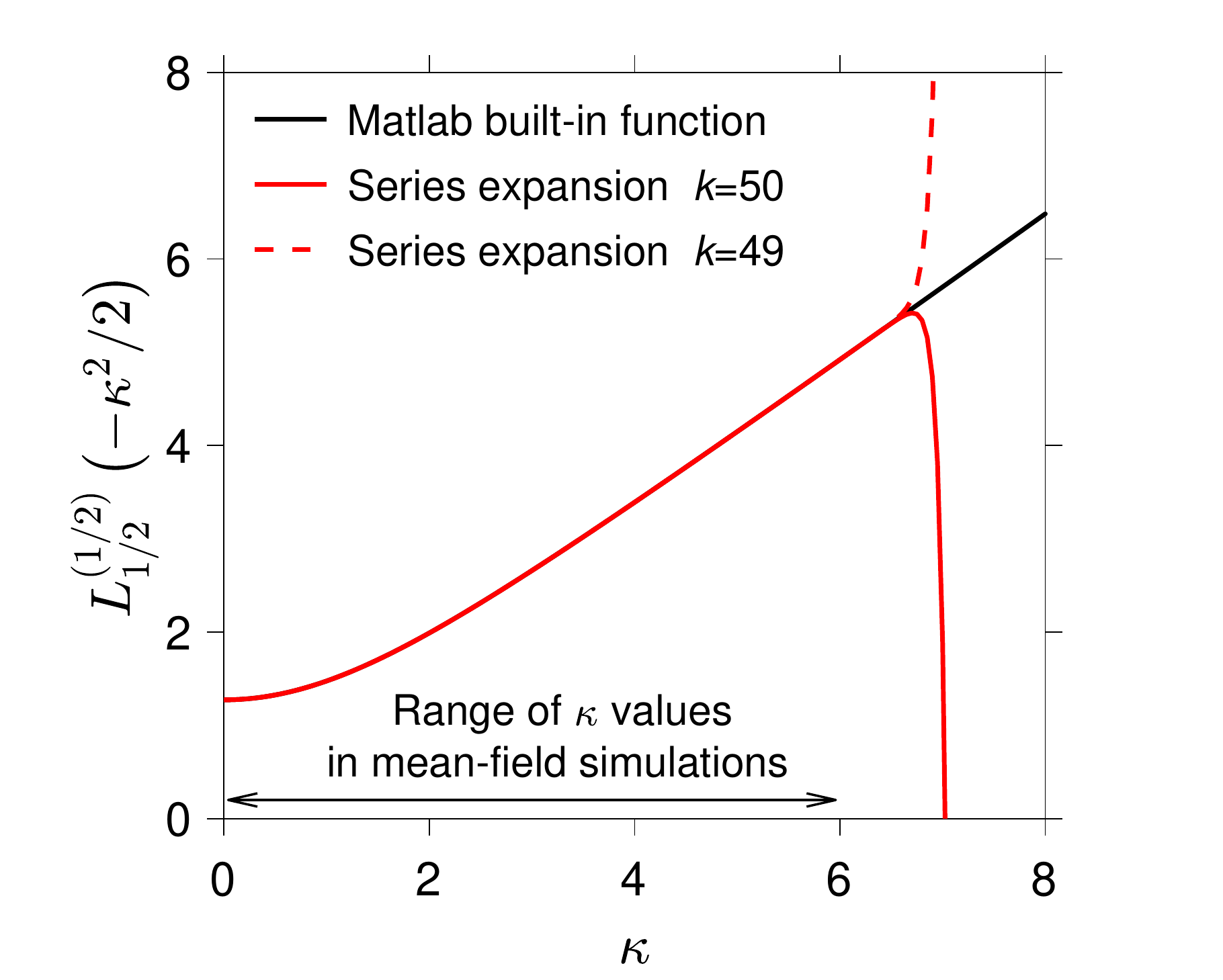}
    \caption{Evolution of $L_{1/2}^{\left(1/2\right)}\left(-\kappa^2/2\right)$ as a function of $\kappa$. The series expansion is compared to the Matlab built-in function.}
    \label{fig:Laguerre}
\end{figure}

In the simulations conducted for this work, the parameter $\kappa$ taken as argument of the Laguerre function remained between 0 and 6. \autoref{fig:Laguerre} shows that the series has converged in the interval [0 6] for $k$ interrupted around 50. Other ways to calculate the function are to use the builtin Laguerre functions existing in standard programming languages, or to use an abacus made from one of the two previous options.

{}

\section{Second derivatives of the energy and mobility laws}
\label{app:2derivative}

The second derivative of the Huang-Humphreys law (\autoref{eq:Huang}) is expressed by:


\begin{equation}
    \mu''\left(\theta\right)=\frac{\mu_c B\eta}{\theta_c^{2\eta}}e^{-B\left(\theta/\theta_c\right)^\eta}
    \left(-B\eta \theta^{2\eta-2}+\theta_c^\eta\left(\eta-1\right)\theta^{\eta-2}\right)
\end{equation}

The second derivative of the Read-Schockley equation (\autoref{eq:ReadSchockley}) is discontinuous at $\theta=\theta_c$. It was choosen to express it as:

\begin{equation}
     \gamma''\left(\theta \right)=  
     \begin{cases} 
    -\frac{\gamma_c}{\theta_c\theta} & \text{if } \theta \leq \theta_c \\
    0 & \text{if } \theta > \theta_c
  \end{cases}
\end{equation}

The effect of this discontinuity on the calculation of boundary energy is negligible since $\gamma''\left(\theta \right)$ already converges towards 0 in its first section.

%
%
%
%
%

\end{appendices}

\end{document}